\documentclass[paper,11pt]{JHEP}
\usepackage[centertags]{amsmath}
\usepackage{amsfonts} \usepackage{amssymb} \usepackage{amsthm}
\allowdisplaybreaks[1]
\usepackage{graphicx}
\newcommand{\Z}{{\mathbb Z}}

\newcommand{\eq}[1]{Eq.~(\ref{#1})}

\newcommand{\ii}{\mathrm{i}}
\newcommand{\ee}{\mathrm{e}}

\newcommand{\lvev}{\big\langle\hskip -3pt\big\langle}
\newcommand{\rvev}{\big\rangle\hskip -3pt\big\rangle}
\newcommand{\vev}[1]{\langle #1 \rangle}

\newcommand{\tr}{\mathrm{tr}\,}

\newcommand{\cA}{{\mathcal{A}}}
\newcommand{\cB}{{\mathcal{B}}}
\newcommand{\cC}{{\mathcal{C}}}

\newcommand{\cE}{{\mathcal{E}}}

\newcommand{\cM}{{\mathcal{M}}}
\newcommand{\cN}{{\mathcal{N}}}
\newcommand{\cO}{{\mathcal{O}}}

\newcommand{\cT}{{\mathcal{T}}}

\newcommand{\cZ}{{\mathcal{Z}}}

\newcommand{\im}{\mbox{Im}\,}
\newcommand{\one}{{\rm 1\kern -.9mm l}}

\newcommand{\Pf}{\mathrm{Pf}}
\newcommand{\cc}{\mathrm{c.c.}}
\title{Holographic non-perturbative corrections 
to gauge couplings
}
\author{M. Bill\'o$^1$, M. Frau$^{1}$, L. Giacone$^{1}$, A. Lerda$^{2}$
\\
\vskip 0.2cm
$^1$ Dipartimento di Fisica Teorica, Universit\`a di Torino\\
and Istituto Nazionale di Fisica Nucleare - sezione di Torino \\
Via P. Giuria 1, I-10125 Torino, Italy\\
\vskip 0.2cm
$^2$Dipartimento di Scienze e Tecnologie Avanzate, Universit\`a del Piemonte
Orientale\\
and I.N.F.N. - Gruppo Collegato di Alessandria - sezione di Torino\\
Viale T. Michel  11, I-15121 Alessandria, Italy\\
\vspace{0.25cm}
\email{billo,frau,giacone,lerda@to.infn.it} 
}
\abstract{We give a direct microscopic derivation of the F-theory background 
that corresponds to four D7 branes of type I$^\prime$ by taking into account the 
D-instanton contributions to the emission of the axio-dilaton field 
in the directions transverse to the D7's. 
The couplings of the axio-dilaton to the D-instanton moduli modify 
its classical source terms which are shown to be
proportional to the elements of the D7 brane chiral ring. 
Solving the bulk field equations with the non-perturbatively corrected sources 
yields the full F-theory background.

This solution represents the gravitational dual of the four-dimensional theory living on 
a probe D3 brane of type I$^\prime$, namely of the $\cN=2$ $\mathrm{Sp}(1)$ SYM theory with $N_f=4$. 
Our results provide an explicit microscopic derivation of the non-perturbative
gravitational dual of this theory. They also explain the recent observation
that the exact coupling for this
theory can be entirely reconstructed from its perturbative part plus the knowledge of 
the chiral ring on the D7 branes supporting its flavor degrees of freedom.
}

\preprint{DFTT/8/2011}

\begin{document}

\section{Introduction}
\label{sec:intro}
The holographic principle finds its incarnation in string theory typically
in the form of a gauge/gravity duality. D branes 
have been crucial in achieving this progress since,
on the one hand, they introduce open string sectors that contain the gauge degrees of
freedom, while on the other hand they source closed string fields and produce
non-trivial space-times.

Matching the realization of the $\mathcal{N}=4$ super Yang-Mills theory by means of open
strings attached to D3 branes in flat space with its closed string description
led Maldacena \cite{Maldacena:1997re} to conjecture the equivalence of this gauge 
theory to type IIB string theory on $\mathrm{AdS}_5\times S_5$; within this 
framework it was soon understood how to holographically relate gauge theory 
correlators to bulk amplitudes \cite{Gubser:1998bc,Witten:1998qj}.
This correspondence has been extended to many other conformal situations and has 
given rise to an entire field of research. 

It is obviously of the greatest interest to study the gauge/gravity duality also
in less supersymmetric and non-conformal cases where the gauge theory couplings
depend on the energy scale. In a stringy description these couplings correspond to dynamical fields from the closed string sector which assume a non-trivial profile in some extra direction dual to the energy scale, along the ideas put forward by Polyakov \cite{Polyakov:1998ju}. The quantum effective couplings should thus be determined by the bulk equations of motion for the corresponding closed string fields, sourced by the brane system on which the gauge theory lives.

These ideas have been exploited, for instance, by constructing non-trivial 
gravitational backgrounds that in some regime can be related to a
system of branes at a conifold singularity \cite{Klebanov:2000hb} 
or to wrapped branes inside a Calabi-Yau manifold \cite{Maldacena:2000yy} 
that support an $\cN=1$ gauge theory in four dimensions. Trusting such 
solutions in the strong coupling regime accounts for many 
expected features of the exact vacua of these $\cN=1$ theories.
Of course, it would be desirable to test the gauge/gravity duality in some
non-conformal case where the exact solution of the gauge theory is known and should be reproduced on the gravitational side, for instance in $\mathcal{N}=2$ gauge theories 
whose exact low-energy description is available after the work of Seiberg and Witten \cite{Seiberg:1994rs,Seiberg:1994aj}.

Non-conformal $\mathcal{N}=2$ $\mathrm{SU}(N)$ SYM theories can be engineered
through fractional D3 branes at a Kleinian singularity (see, {\it e.g.} Ref.s~\cite{Klebanov:1999rd}\nocite{Bertolini:2000dk,Polchinski:2000mx,Bertolini:2001qa} - \cite{Billo:2001vg}). 
In this set-up the gauge coupling is typically represented by a scalar field from the closed string twisted sector for which the D3 branes at the singularity act as $\delta$-function sources localized
in the remaining two transverse directions.
Solving the corresponding field equations yields a logarithmic profile that exactly matches the perturbative running of the coupling constant upon identifying the two transverse directions 
with the (complexified) scale.
These perturbative checks have been successfully extended to various $\mathcal{N}=2$ and
$\mathcal{N}=1$ theories realized in orbifold set-ups \cite{DiVecchia:2005vm}.

However, the exact gauge coupling generically contains also a series
of non-perturbative corrections. Are these reproduced on the gravitational 
side? And how? Clearly, the logarithmic solution described above has to be modified, as 
a consequence also of the back-reaction of the branes on the geometry of the system.
In the $\mathcal{N}=2$ case it has been speculated that in the large-$N$ limit the correct modification 
is through an ``enhan\c{c}on'' mechanism \cite{Johnson:1999qt} in which the D3 branes actually 
expand in the transverse space, forming a ring of radius proportional to the dynamically generated 
scale where non-perturbative effects become relevant. The resulting profile of the coupling field 
agrees with the SW solution in this limit \cite{Petrini:2001fk},
but this mechanism lacks an explicit microscopic derivation.

For finite $N$, the non-perturbative corrections to the gauge coupling contained in the SW solution have been retrieved via multi-instanton calculus and localization techniques by Nekrasov \cite{Nekrasov:2002qd,Nekrasov:2003rj}. In the stringy description, they are provided by the inclusion of D-instanton sectors in addition to the 
the D3/D3 open strings \cite{Green:2000ke,Billo:2002hm}. 
It is therefore natural to speculate that the corresponding non-perturbative modifications of the dual gravitational solution be provided again by D-instantons
which indeed modify closed string interactions. This was in fact 
the focus of the early interest in these objects. 
In particular, in Ref.s~\cite{Green:1997tv}\nocite{Green:1997tn} - \cite{Green:1998yf} the key ideas 
and techniques to elucidate the D-instanton contributions to the gravitational 
effective action were developed. In the meantime, 
it was recognized that D-instantons represent the stringy counterpart of gauge 
instantons \cite{Witten:1995gx,Douglas:1995bn} and the investigation of 
their effects in the open string sector was pursued in many 
directions.
A key step in extracting multi-instanton corrections is the problematic integration 
over the moduli space. 
In supersymmetric cases, though, localization techniques \cite{Moore:1998et,Nekrasov:2002qd,Nekrasov:2003rj} have allowed to overcome 
this difficulty both in the field-theory \cite{Bruzzo:2002xf,Flume:2002az} and in the
string-theory description \cite{Billo:2009di}\nocite{Fucito:2009rs} - \cite{Billo':2010bd}.
Of course, the integration over moduli space is crucial also to determine
D-instanton effects in the closed string sector
and being able to rely on localization techniques will be thus extremely
useful for this purpose.. 

We expect that D-instantons affect the equation of motion for the closed string 
field that represents the gauge coupling by modifying the source with non-perturbative terms, 
so that the new profile of this field coincides with the exact coupling in the gauge theory. 
We set out to check this expectation by considering a particular case where the $\mathcal{N}=2$ gauge 
theory is $\mathrm{SU}(2)$ with $N_f=4$. This theory has vanishing $\beta$-function but, 
when masses for the fundamental hypermultiplets are turned on, the 
effective coupling at low energies receives non-perturbative corrections; its 
exact expression is contained in the SW curve for this model 
\cite{Seiberg:1994aj} and has been recently worked out in Ref.~\cite{Billo:2010mg}. 
Accounting for this exact coupling from a dual 
gravitational perspective would represent a valuable step forward. 

This conformal theory can be realized on the world-volume of a
D3 brane in a local version of type I$^\prime$ superstring theory. The D3 brane
(together with its orientifold image) supports an $\cN=2$
$\mathrm{Sp}(1)\!\sim\!\mathrm{SU}(2)$ gauge theory, with fundamental hypermultiplets
provided by the open strings stretching between the D3 brane and the four
D7 branes (plus the orientifold O7 plane) sitting near one orientifold fixed
point. The gauge coupling corresponds to the axio-dilaton field; finding its
exact expression amounts thus to determine the consistent F-theory background
for this set-up.

This task was tackled long ago by Sen \cite{Sen:1996vd} who noticed that
the na\"ive axio-dilaton profile produced by the D7 branes and the orientifold,
has logarithmic singularities at the sources' locations. However, such singularities 
are incompatible with the physical interpretation of the dilaton as the string coupling, so that
the exact profile must be modified. Based on the symmetries of the problem, Sen
proposed that the F-theory axio-dilaton coincides with the
exact gauge coupling encoded in the SW solution for the $\cN=2$ SU(2) SYM theory
with $N_f=4$; this connection was later explained \cite{Banks:1996nj} in terms of a D3 brane probing this background. Recently \cite{Billo:2010mg}, it has been shown
that the exact SW solution is retrieved by including D-instanton corrections 
in the D7/O7/D3 brane system. In this chain of arguments, the
gauge/gravity relation is assumed and exploited to express the gravitational
background in terms of the known solution of the dual gauge theory. Our purpose
is to show that it is possible to compute directly the non-perturbative
completion of the gravitational background, and in particular of the axio-dilaton
profile, without reference to the gauge theory; getting the same expression of
the exact gauge coupling amounts then to a non-trivial check of the
gauge/gravity duality at the non-perturbative level.

In this paper, therefore, the D3 brane supporting the $\cN=2$ SYM theory will play no
r\^ole. We will include D-instanton corrections to the local type I$^\prime$ system
of D7 branes plus orientifold, and show how the exact F-theory background
emerges in this way. We think this is important at the conceptual level and
believe that the techniques we develop here to handle the non-perturbative
corrections to the profile of fields from the closed string sector
could be useful in many situations. 

In Section~\ref{sec:dcd7} we set the stage by considering the effect of
displacing the D7 branes from the orientifold by giving a classical expectation
value to the adjoint scalar field living on their world-volume and explore the
axio-dilaton couplings to the D7 world-volume; this determines how the
displaced D7 brane can source this field. In Section~\ref{sec:dcmoduli} we add
D(--1) branes to the system and consider the couplings of the instanton moduli
to the axio-dilaton field. These couplings arise from disks having their
boundary attached to the D-instantons with a closed string axio-dilaton vertex
in the interior and multiple moduli vertex insertions on the boundary;
all diagrams relevant for our purposes turn out (rather surprisingly) to be
computable. 

We then explain how to extract the non-perturbative corrections to the
axio-dilaton sources by suitably integrating these
couplings over the instanton moduli space. 
We demonstrate in Section~\ref{sec:profile} that the relevant 
integrals coincide with the ones that compute the non-perturbative
corrections to the elements of the chiral ring on the D7 branes, namely to the
vacuum expectation values of the traces of the powers of the adjoint D7/D7 scalar; 
these integrals can be computed by means of localization techniques \cite{Fucito:2009rs}. Thus
we explicitly show that the non-perturbative modifications of the source terms
replace powers of the classical expectation values of the D7 brane adjoint
scalar with the corresponding non-perturbative quantum vacuum expectation values, a fact that we
already observed (without proving it) in Ref.~\cite{Billo:2010mg}. Finally, to show
that our techniques can be applied in different contexts, and in particular do
not require an eight-dimensional sector, we consider in Section
\ref{sec:orbifold} a slight modification of our set-up in which a $\Z_2$
orbifold projection in enforced in four of the D7 brane world-volume directions.
The computation of the axio-dilaton couplings to the D(--1) moduli goes through
with suitable modifications (and is in fact slightly simpler), so that also in
this case it is possible to obtain the non-perturbative source terms and hence
the exact axio-dilaton profile.

The three Appendices contain our notations and conventions, technical details and explicit 
computations that would not fit conveniently in the main text.

\section{Axio-dilaton couplings of the D7 action}
\label{sec:dcd7}
\subsection{The set-up}
\label{subsec:setup}
We consider the so-called type I$^\prime$ superstring theory,
that is the projection of the type IIB theory by
\begin{equation}
\label{typeipd}
\Omega=\omega\, \mathbf{I}_2\, (-1)^{F_L}~,
\end{equation}
where $\omega$ is the world-sheet orientation reversal, $\mathbf{I}_2$ is the
inversion of two coordinates, say $x^{8,9} \to - x^{8,9}$, and $F_L$ is the
target-space left-moving fermion number. If the last two directions are
compactified on a torus $\cT_2$, the type I$^\prime$ model is T-dual on this torus to the standard
type I theory and possesses four O7 planes located at the four fixed points of
$\cT_2$. Tadpole cancellation requires the presence of sixteen D7 branes (plus their
images); if the D7 branes are distributed in groups of four and placed on top of
the O7 planes, the cancellation is local and there is no backreaction on the
geometry; in particular, the axio-dilaton field 
\begin{equation}
\tau = C_0 + \ii\, \ee^{-\phi}
\label{dilaxio}
\end{equation}
is constant also along the directions of $\cT_2$, which we will parametrize by
the complex coordinate $z =x^8 -\ii x^9$.

In the following, we will consider a ``local'' limit around one of the $\cT_2$
fixed points, say $z=0$, taking thus the transverse space to be simply
$\mathbb{C}$, and we will consider moving the four D7 branes out of this fixed point.
In this situation, local tadpole cancellation is lost and the axio-dilaton
profile becomes non-trivial. We split the field $\tau$ into its expectation value 
$\tau_0 = \ii/g_s$, representing the inverse string coupling, and a fluctuation part 
$\widetilde\tau$, namely
\begin{equation}
 \tau = \tau_0 + \widetilde\tau
\label{tauspli}
\end{equation}
with
\begin{equation}
\label{fluc}
\widetilde\tau = \widetilde C_0 + \frac{\ii}{g_s}\big(\ee^{-\widetilde\phi} -1\big)\sim
\widetilde C_0 -\frac{\ii}{g_s}\,\widetilde\phi~,
\end{equation}
where in the last step we retained only the linear term in the fluctuations.
The bulk kinetic term for this field reads%
\footnote{It arises from the term
\begin{equation*}
\label{bulkda}
-\frac{1}{2\kappa^2} \int d^{10}x~ \frac{\partial_M \overline\tau
\,\partial^M\tau}{(\im \tau)^2}~,
\end{equation*} 
in the Einstein frame supergravity action, where $\kappa =g_s\widetilde\kappa$.}
\begin{equation}
\label{bulk}
S_{\mathrm{bulk}} = -\frac{1}{2\widetilde\kappa^{\,2}} \int d^{10}x~ \partial_M
\widetilde{\overline\tau}\, \partial^M\widetilde \tau~,
\end{equation} 
where $\widetilde{\overline\tau}$ is the complex
conjugate of $\widetilde{\tau}$ and $\widetilde\kappa = 8\pi^{\frac 72} \alpha'^{2}$.
Varying in $\widetilde{\overline\tau}$, the bulk contribution to the field equation 
is thus proportional to $\square\,\widetilde\tau$.

As we will review in the next sub-section, the O7 plane and the D7 branes act as sources
for the axio-dilaton localized in the two common transverse directions, thus leading to a
logarithmic dependence in $z$. However, this behavior is not acceptable, since
the imaginary part of $\tau$, representing the inverse string coupling constant,
blows up at the sources' locations. This behavior is in fact modified non-perturbatively, and
the correct background corresponds to the particular limit of F-theory
considered long ago by Sen in Ref.~\cite{Sen:1996vd}. Here we will show explicitly
how the non-perturbative corrections to the axio-dilaton profile are induced by
D-instantons.

To do this, we will need to carefully consider the couplings of $\tau$ to the D7
branes and to the D(--1) moduli. A crucial r\^ole in this analysis is played by
the fact that the axio-dilaton is the lowest component
of a chiral superfield $T$ which contains the massless closed string degrees
of freedom of the type I$^\prime$ theory%
\footnote{In the context of type IIB superstring in $d=10$, the analogous organization
of the supergravity degrees of freedom in an analytic superfield with lowest
component the axio-dilaton \cite{Schwarz:1983qr}\nocite{Howe:1983sra,deHaro:2002vk} - \cite{Green:2003an} represented a key point in the investigation of the D-instanton effects 
on the effective action carried out in
\cite{Green:1997tv,Green:2000ke}.
We will make use of many of the ideas and techniques of these papers but, 
since we are interested in the couplings of the axio-dilaton itself, for us the relevant 
component of the superfield will be the highest one, rather than some of the lower components 
containing different physical fields.}
\begin{equation}
\label{Tsuper}
T = \tau_0 + \widetilde T = \tau_0 +\widetilde\tau + \sqrt{2}\theta \widetilde\lambda + ... + 2 \theta^8 \frac{\partial^4~}{\partial z^4}\widetilde{\overline\tau}~. 
\end{equation} 
Here $\widetilde\lambda$ is the dilatino, while the remaining supergravity degrees of freedom 
appear in the omitted terms in the $\theta$-expansion. Our notation (see Appendix
\ref{app:notconv}) is that the sixteen supersymmetries of type I$^\prime$ can be arranged in
a Majorana-Weyl $10d$ spinor $\Theta^\cA$, which in turn, under the $10\to 8 + 2$
split of target space intrinsic to the theory, decomposes into a chiral and an
anti-chiral $8d$ spinor: $\Theta^\cA\to
(\theta^{\alpha},\bar\theta^{\dot\alpha})$. 
For our purposes the most important component of the $T$ superfield is the highest
one, proportional to $\theta^8$. Its expression involves the fourth holomorphic
derivative of the complex conjugate of $\widetilde\tau$, as we will check explicitly in
Appendix \ref{subapp:C0theta} by the computation of a disk diagram with eight
$\theta$ insertions.

The type I$^\prime$ theory contains also an open string sector, made by strings whose
endpoints are stuck on the D7 branes. The orientifold projection implies that the
massless content of this sector is that of an $8d$ gauge theory.
In our local model, the gauge group of this theory is SO(8) and the massless open string degrees of
freedom can be arranged in a chiral superfield
\begin{equation}
\label{defM}
M = m + \sqrt{2} \,\theta \Lambda + \frac 12 \,\theta \gamma^{\mu\nu} \theta\,
F_{\mu\nu} + \ldots~.
\end{equation}
Here $m$ is a complex scalar, $\Lambda$ is the gaugino and $F_{\mu\nu}$ is the gauge field strength, 
all in the adjoint representation of SO(8).

\subsection{Tree level couplings}
\label{treecd7}
Let us now consider the tree-level linear coupling of the axio-dilaton field to the
D7 branes. 

\paragraph{D7 branes at the origin}

As well known \cite{Polchinski:1995mt}, D$p$ branes introduce tadpoles for closed string
fields, realized by disks whose boundary lies on the D branes and with a closed
string vertex in the interior. These tadpoles can be evaluated taking the inner
product of the boundary states representing the branes with the relevant closed
string states \cite{Polchinski:1995mt,DiVecchia:1997pr}. Let us momentarily generalize our
set-up to the case of $N_f$ D7 branes (instead of four) and an O7 plane placed
at the origin of the transverse space spanned by $z$.
The boundary state of the D7 branes, as well as the crosscap state representing the O7
plane, couples to the fluctuations of the dilaton and of the Ramond-Ramond eight-form $C_8$,
which is related to $C_0$ by bulk Poincar\'e duality  
\begin{equation}
\label{C8C0}
dC_8 = * dC_0~.
\end{equation} 
The profile of these fields can be obtained from the emission amplitude in momentum space,
stripped of the polarization, by attaching a free propagator 
and taking the Fourier transform \cite{DiVecchia:1997pr}. 

Equivalently, the linear couplings of the D7 branes to the dilaton and axion
fluctuations are encoded in their world-volume theory, whose well-known
structure is worth recalling briefly here. In the Einstein frame
the tree-level Born-Infeld action for $N_f$ D7 branes (and their orientifold images) 
takes the schematic form
\begin{equation}
\label{bi}
S_{\mathrm{tree}}^{\mathrm{BI}} 
=-\frac{T_7}{\kappa} \int_{\mathrm{D7}} \!d^8x\, \left\{2N_f\,\ee^{\widetilde\phi} - 
\frac{(2\pi\alpha')^2}{4}\,\tr F^2 + \frac{(2\pi\alpha')^4}{12}\,\ee^{-\widetilde\phi} \,
\tr(t_8 F^4) + O(F^5)\right\}~.
\end{equation}
Here tensor $t_8$ is the anti-symmetric eight-index tensor appearing in various 
superstring amplitudes (see {\it e.g.} Ref.~\cite{Green:1987mn}),
and $T_7 = \sqrt{\pi}(2\pi\sqrt{\alpha'})^{-4}$ is the D7 brane tension.

The linear part in $\widetilde\phi$ in the first term of \eq{bi} corresponds to
the coupling of the dilaton fluctuation to the D7 boundary state, and represents
a source localized on the D7 branes given by
\begin{equation}
\label{st}
-2 N_f \frac{T_7}{\kappa}\int \!d^{10}x~\widetilde\phi ~\delta^2(z)~.
\end{equation}
Taking into account the bulk kinetic term (\ref{bulk}), and using
the fact that $2 T_7\kappa = g_s$, the field equation for $\widetilde\phi$ reads
\begin{equation}
\label{eomphi}
\square\,\widetilde\phi = 2 g_s N_f\, \delta^2(z)~,
\end{equation}
and its solution is $\widetilde\phi = (1/\pi)N_f g_s \log |z|$. If we include
the O7 source term, that possesses a negative charge and shifts $N_f$ to
$(N_f -4)$ and properly add the axion $\widetilde C_0$ induced by the sources for its 
Poincar\'e dual 8-form, we obtain the classical profile for $\widetilde\tau$ 
which has the following logarithmic behavior 
\begin{equation}
\label{resemd7}
\widetilde\tau_{\mathrm{cl}}(z) = \frac{1}{2\pi\ii} (2 N_f - 8) \log
\frac{z}{z_0} 
\end{equation}
where $z_0$ is a suitable length scale. For $N_f=4$, $\tau$ is constant since
the D7 and O7 charges cancel locally.

The remaining terms in the action (\ref{bi}) describe the interactions of
the dilaton with the gauge field. Note that there is no
coupling to the quadratic Yang-Mills Lagrangian, while the dilaton fluctuation couples to
the quartic terms in $F$. If we take into account also the Wess-Zumino action, 
these quartic terms read
\begin{equation}
\label{quarticgf}
S_{\mathrm{tree}}^{(4)} = -\frac{1}{96\pi^3\, g_s} \int_{\mathrm{D7}} \!d^8x\, \ee^{-\widetilde\phi}\,
\tr(t_8 F^4)-\frac{\ii}{192\pi^3} \int_{\mathrm{D7}} 
\!\widetilde C_0 \,\tr(F\wedge F\wedge F\wedge F)~. 
\end{equation}
Using the superfield $M$ introduced in \eq{defM}, they can be rewritten 
as a superpotential term
\begin{equation}
\label{4prep}
S_{\mathrm{tree}}^{(4)}
= \frac{1}{(2\pi)^4} \int d^8x\, d^8\theta\,
F_{\mathrm{tree}}^{(4)}
+ \mathrm{c.c.}~,
\end{equation}
with
\begin{equation}
\label{quartict}
F_{\mathrm{tree}}^{(4)}
= \frac{\ii\pi}{12}\,\tau\, \tr M^4~.
\end{equation}
If the SO(8) gauge field gets a non-zero classical expectation value,
\eq{quarticgf} may give rise to source terms for $\widetilde\phi$ and $\widetilde C_0$. 
For instance, a constant background field $F$ yields a source term for $\widetilde\phi$
proportional to $\tr(t_8 F^4)$. On the other hand, in topologically non-trivial sectors 
with fourth Chern number $k$, the vacuum is represented by instanton-like configurations%
\footnote{As discussed in Ref.~\cite{Billo':2009gc}, for $k=1$ such a
configuration corresponds to the point-like limit of the so-called SO(8)
instanton \cite{Grossman:1989bb,Grossman:1984pi}.} 
that minimize the action (\ref{quarticgf}) (at fixed axio-dilaton)
yielding simply 
$S_{\mathrm{tree}}^{(4)} = -{2\pi\ii}\,k\,\tau$.
This is the same action that describes the coupling to the axio-dilaton
of $k$ D-instantons, which indeed represent these configurations 
in the string picture \cite{Billo':2009gc,Billo:2009di}.

The message we obtain from this analysis is two-fold: on the one side, this suggests that at the
non-perturbative level we must take into account the interactions of the
axio-dilaton with the D-instantons. On the other side, we see that extra source
terms for the axio-dilaton can arise from its interactions with the open string
degrees of freedom, if the latter acquire non-trivial expectation values. Such
interaction terms are not limited to the one usually considered in the Born-Infeld and Wess-Zumino
actions but include other structures, as we will now discuss.

\paragraph{Displaced D7 branes and scalar field couplings}

If we modify our set-up by placing the D7 branes at the positions $\pm z_i$,
the classical axio-dilaton profile becomes
\begin{equation}
\label{massiveprofile}
\widetilde\tau_{\mathrm{cl}}(z) = \frac{1}{2\pi\ii} \Bigl\{\sum_{i=1}^{N_f} 
\Bigl[\log \frac{z -z_i}{z_0}
+ \log \frac{z + z_i}{z_0}\Bigr] - 8 \log \frac{z}{z_0} \Bigr\}~.
\end{equation}
This is non-trivial even in the case $N_f=4$.

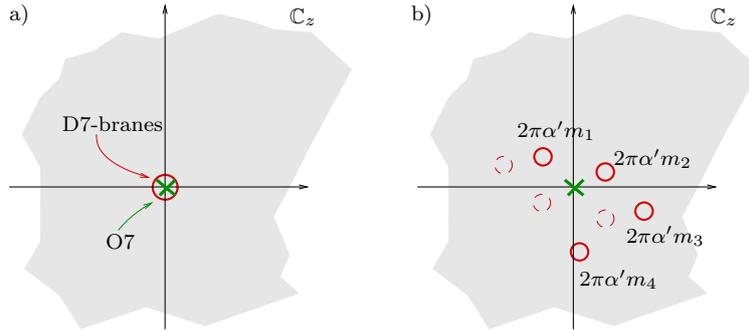
\begin{figure}
\begin{center}
\begin{picture}(0,0)%
\includegraphics{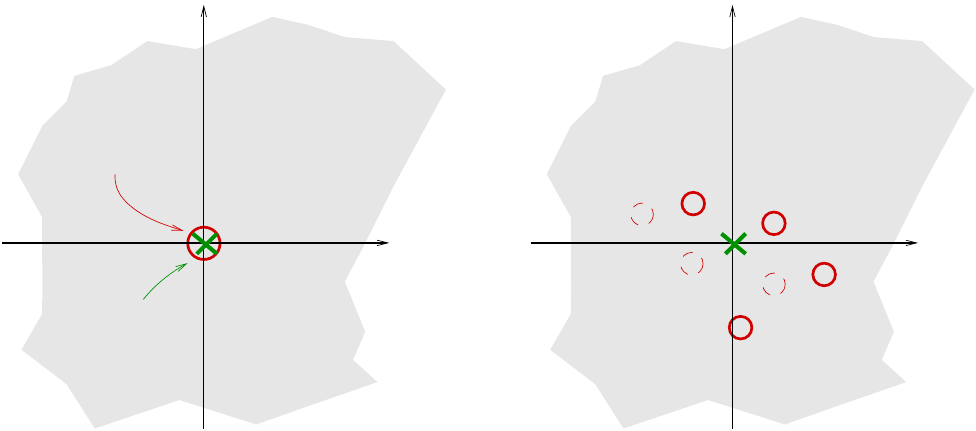}%
\end{picture}%
\setlength{\unitlength}{1699sp}%
\begingroup\makeatletter\ifx\SetFigFont\undefined%
\gdef\SetFigFont#1#2#3#4#5{%
  \reset@font\fontsize{#1}{#2pt}%
  \fontfamily{#3}\fontseries{#4}\fontshape{#5}%
  \selectfont}%
\fi\endgroup%
\begin{picture}(10868,4769)(249,-4178)
\put(1666,-2986){\makebox(0,0)[lb]{\smash{{\SetFigFont{8}{9.6}{\familydefault}{\mddefault}{\updefault}O7}}}}
\put(1036,-1276){\makebox(0,0)[lb]{\smash{{\SetFigFont{8}{9.6}{\familydefault}{\mddefault}{\updefault}D7-branes}}}}
\put(271,344){\makebox(0,0)[lb]{\smash{{\SetFigFont{8}{9.6}{\familydefault}{\mddefault}{\updefault}a)}}}}
\put(6076,344){\makebox(0,0)[lb]{\smash{{\SetFigFont{8}{9.6}{\familydefault}{\mddefault}{\updefault}b)}}}}
\put(7606,-1366){\makebox(0,0)[lb]{\smash{{\SetFigFont{8}{9.6}{\familydefault}{\mddefault}{\updefault}$2\pi\alpha' m_1$}}}}
\put(9001,-1771){\makebox(0,0)[lb]{\smash{{\SetFigFont{8}{9.6}{\familydefault}{\mddefault}{\updefault}$2\pi\alpha' m_2$}}}}
\put(8506,-3526){\makebox(0,0)[lb]{\smash{{\SetFigFont{8}{9.6}{\familydefault}{\mddefault}{\updefault}$2\pi\alpha' m_4$}}}}
\put(9181,-2896){\makebox(0,0)[lb]{\smash{{\SetFigFont{8}{9.6}{\familydefault}{\mddefault}{\updefault}$2\pi\alpha' m_3$}}}}
\put(4321,299){\makebox(0,0)[lb]{\smash{{\SetFigFont{8}{9.6}{\familydefault}{\mddefault}{\updefault}$\mathbb{C}_z$}}}}
\put(10396,299){\makebox(0,0)[lb]{\smash{{\SetFigFont{8}{9.6}{\familydefault}{\mddefault}{\updefault}$\mathbb{C}_z$}}}}
\end{picture}%
\end{center}
\caption{a) The D7 branes and the O7 plane are placed at the origin. b) The
D7 branes (and their images) are displaced from the origin; this corresponds to
distribution of charges which leads to the profile given in Eq.s
(\ref{massiveprofile})-(\ref{massexpm}) for the axio-dilaton.}
\label{fig:displace}
\end{figure}

In the following we concentrate on this case. The dependence from the scale
$z_0$ disappears and we can expand the profile (\ref{massiveprofile}) for large
$z$, getting
\begin{equation}
\label{massexp}
\widetilde\tau_{\mathrm{cl}}(z) = - \frac{1}{2\pi\ii}\sum_{\ell=1}^\infty
\frac{1}{\ell}\frac{\sum_{i=1}^4 z_i^{2\ell}}{z^{2\ell}}~. 
\end{equation} 
Displacing the D7 branes corresponds to giving a classical expectation value
\begin{equation}
\label{vevm}
m_{\mathrm{cl}} = \mathrm{diag}\{m_1,m_2,m_3,m_4,-m_1,-m_2,-m_3,-m_4\}~~~~~\mbox{with}~ m_i =
\frac{z_i}{2\pi\alpha'}~,
\end{equation}
to the SO$(8)$ adjoint scalar field $m$ of the D7 brane world-volume theory,
which has canonical dimension of $\mathrm{(length)}^{-1}$. In terms of $m_{\mathrm{cl}}$, the
profile (\ref{massexp}) reads
\begin{equation}
\label{massexpm}
\widetilde\tau_{\mathrm{cl}}(z) = - \frac{1}{2\pi\ii}\sum_{\ell=1}^\infty
\frac{(2\pi\alpha')^{2\ell}}{2\ell}\,\frac{\tr m_{\mathrm{cl}}^{2\ell}}{z^{2\ell}}~,
\end{equation} 
which solves the following differential equation
\begin{equation}
\label{eomtaub}
\square\,\widetilde\tau = -2\ii \sum_{\ell=1}^\infty \frac{(2\pi\alpha')^{2\ell}\,\tr
m_{\mathrm{cl}}^{2\ell}}{(2\ell)!} \, \frac{\partial^{2\ell}\delta^{2}(z)}{\partial
z^{2\ell}}~.
\end{equation}
This is the field equation obtained by varying with respect to $\widetilde{\overline\tau}$
an action which, in addition to the bulk term (\ref{bulk}), contains also a
source term localized on the world-volume of the D7 branes
\begin{equation}
\label{sbst}
S_{\mathrm{source}} = - \frac{T_7}{\widetilde\kappa} \int_{\mathrm{D7}} \!d^8x \, J_{\mathrm{cl}}\,
\widetilde{\overline\tau} + \mathrm{c.c}~.
\end{equation}
Indeed, requiring that
\begin{equation}
\label{variaztb}
\frac{\delta\phantom{\widetilde{\overline\tau}}}{\delta\widetilde{\overline\tau}}\left(S_{\mathrm{bulk}}
+ S_{\mathrm{source}}\right) =0~,
\end{equation}
we obtain 
\begin{equation}
\label{eomtb}
\square\,\widetilde\tau = J_{\mathrm{cl}}\, \delta^{2}(z)~,
\end{equation}
which coincides with \eq{eomtaub} if 
\begin{equation}
\label{Jpis}
J_{\mathrm{cl}} = - 2 \ii  \sum_{\ell=1}^\infty 
\frac{(2\pi\alpha')^{2\ell}\,\tr m_{\mathrm{cl}}^{2\ell} }{(2\ell)!} 
\frac{\partial^{2\ell}}{\partial z^{2\ell}}~.
\end{equation}
In momentum space, this current becomes
\begin{equation}
\label{Jft}
J_{\mathrm{cl}} = - 2 \ii  \sum_{\ell=1}^\infty (-1)^\ell
\frac{(2\pi\alpha')^{2\ell}\,\tr m_{\mathrm{cl}}^{2\ell} }{(2\ell)!} \,{\bar p}^{2\ell}~, 
\end{equation}
where $\bar p = (p_8 + \ii p_9)/2$ is the momentum conjugate to $z$.

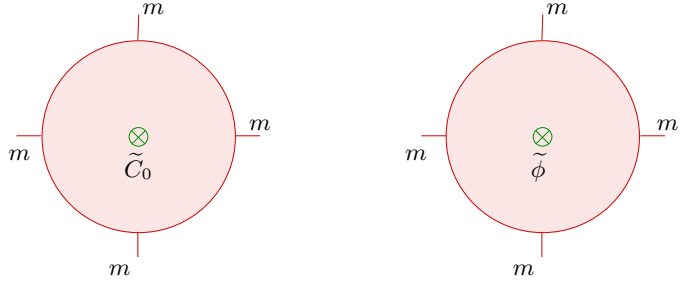
\begin{figure}
\begin{center}
\begin{picture}(0,0)%
\includegraphics{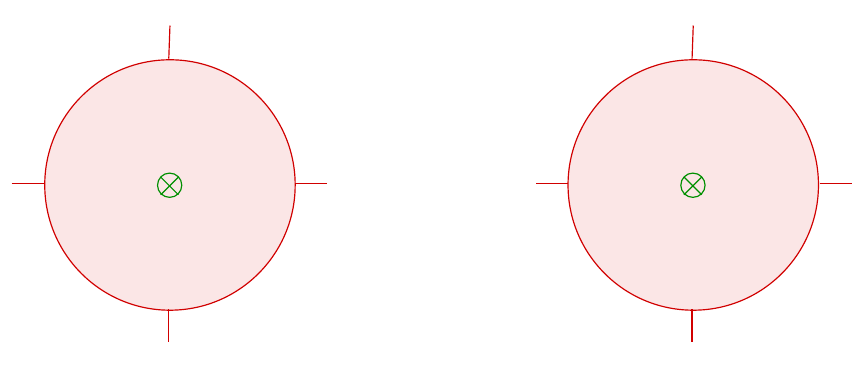}%
\end{picture}%
\setlength{\unitlength}{1492sp}%
\begingroup\makeatletter\ifx\SetFigFont\undefined%
\gdef\SetFigFont#1#2#3#4#5{%
  \reset@font\fontsize{#1}{#2pt}%
  \fontfamily{#3}\fontseries{#4}\fontshape{#5}%
  \selectfont}%
\fi\endgroup%
\begin{picture}(10837,4744)(-14,-4289)
\put(8296,-4171){\makebox(0,0)[lb]{\smash{{\SetFigFont{9}{10.8}{\familydefault}{\mddefault}{\updefault}$m$}}}}
\put(8851,164){\makebox(0,0)[lb]{\smash{{\SetFigFont{9}{10.8}{\familydefault}{\mddefault}{\updefault}$m$}}}}
\put(8591,-2536){\makebox(0,0)[lb]{\smash{{\SetFigFont{9}{10.8}{\familydefault}{\mddefault}{\updefault}${\widetilde \phi}$}}}}
\put(1651,-4171){\makebox(0,0)[lb]{\smash{{\SetFigFont{9}{10.8}{\familydefault}{\mddefault}{\updefault}$m$}}}}
\put(2206,164){\makebox(0,0)[lb]{\smash{{\SetFigFont{9}{10.8}{\familydefault}{\mddefault}{\updefault}$m$}}}}
\put(1906,-2536){\makebox(0,0)[lb]{\smash{{\SetFigFont{9}{10.8}{\familydefault}{\mddefault}{\updefault}${\widetilde C_0}$}}}}
\put(3961,-1771){\makebox(0,0)[lb]{\smash{{\SetFigFont{9}{10.8}{\familydefault}{\mddefault}{\updefault}$m$}}}}
\put(  1,-2266){\makebox(0,0)[lb]{\smash{{\SetFigFont{9}{10.8}{\familydefault}{\mddefault}{\updefault}$m$}}}}
\put(10666,-1771){\makebox(0,0)[lb]{\smash{{\SetFigFont{9}{10.8}{\familydefault}{\mddefault}{\updefault}$m$}}}}
\put(6526,-2266){\makebox(0,0)[lb]{\smash{{\SetFigFont{9}{10.8}{\familydefault}{\mddefault}{\updefault}$m$}}}}
\end{picture}%
\end{center}
\caption{The diagrams describing the coupling of the Ramond-Ramond 0-form and dilaton
fluctuations to four scalar fields $m$ of the D7 brane theory. These amplitudes
in momentum space turn out to be proportional, as shown in Appendix
\ref{app:sd}, to $\tr m^4\, (\bar p)^4$, which leads, upon insertion of the free
propagator and Fourier transform, to a profile $\sim \tr m^4/z^4$.}
\label{fig:mmcphi}
\end{figure}

{From} a diagrammatic point of view, the source terms (\ref{Jft}) arise from
emission diagrams for $\tau$ from disks with multiple insertions of $m$ (which
turn out to be proportional to powers of the transverse momentum $\bar p$);
an example is depicted in Fig.~\ref{fig:mmcphi}. This means that the D7 action 
must contain interaction
terms linear in $\widetilde{\overline\tau}$ of the form (see \eq{sbst})
\begin{equation}
\label{smtau}
\begin{aligned}
& 2\ii\,\frac{T_7}{\widetilde\kappa} \int_{\mathrm{D7}}\! d^8x\,
\sum_{\ell=1}^\infty \frac{(2\pi\alpha')^{2\ell}\,\tr m^{2\ell} }{(2\ell)!}
\frac{\partial^{2\ell} \,\widetilde{\overline\tau}}{\partial z^{2\ell}} + \cc
\\
&~=  \frac{4\pi\ii}{(2\pi)^4} \int_{\mathrm{D7}}\! d^8x\,
\sum_{\ell=1}^\infty \frac{(2\pi\alpha')^{2\ell-4}\,\tr m^{2\ell} }{(2\ell)!}
\frac{\partial^{2\ell} \,\widetilde{\overline\tau}}{\partial z^{2\ell}} + \cc~,
\end{aligned}
\end{equation}
which, by freezing $m$ to its expectation value $m_{\mathrm{cl}}$, reproduce precisely the source
terms of \eq{Jft}. We now show that these new terms can be easily incorporated by generalizing
the previous analysis.

\paragraph{Superpotential contributions}

Starting from the terms quartic in the adjoint scalar $m$, these interactions, together with other
terms related by supersymmetry, can be expressed as contributions to an 
$8d$ ``superpotential'', using the chiral bulk superfield $T$ of \eq{Tsuper} 
and of the open string superfield $M$ of \eq{defM}. 
Indeed, the $\ell=2$ term in \eq{smtau} is contained in the action
\begin{equation}
\label{4prepT}
S_{\mathrm{tree}}^{(4)} = \frac{1}{(2\pi)^4} \int\! d^8x\, d^8\theta\,
F_{\mathrm{tree}}^{(4)} + \mathrm{c.c.}~,
\end{equation}
where the superpotential is given by
\begin{equation}
\label{quarticT}
F_{\mathrm{tree}}^{(4)} = \frac{2\pi\ii}{4!}\, \tr M^4\, T~,
\end{equation}
and depends on $x$ and $\theta$ through the superfields $M$ and $T$.
This is just the expression (\ref{quartict}) that accounts for the quartic terms
in the gauge field in which, however, the axio-dilaton $\tau$ has been promoted 
to the corresponding superfield $T$. Notice that in \eq{4prepT}
we can saturate the integration over $d^8\theta$ in different ways. 
If we pick up all eight $\theta$'s from the $M^4$ factor, we retrieve the quartic 
gauge action (\ref{quarticgf}). At the opposite end, we can take all the eight $\theta$'s from
the $T$ superfield and, recalling the expansion (\ref{Tsuper}), we obtain
\begin{equation}
\label{qit}
\frac{4\pi\ii}{(2\pi)^4} \int_{\mathrm{D7}}\! d^8x\,
\frac{\tr m^{4} }{4!}
\frac{\partial^{4} \,\widetilde{\overline\tau}}{\partial z^{4}} + \cc~,
\end{equation}
which coincides with the $\ell=2$ term in \eq{smtau}.
The terms with higher values of $\ell$ can similarly be written as superpotential terms%
\footnote{Also the term quadratic in the scalar field
$m$, corresponding to $\ell=1$, can actually be written as a superpotential
contribution, at the price of allowing for a non-locality in the transverse
directions: \begin{equation*} \label{quadterm} F_{\mathrm{tree}}^{(2)} =
\frac{2\pi\ii}{(2\pi\alpha')^2} \frac{\tr M^2}{2}
\left(\frac{\partial~}{\partial z}\right)^{-2} T~. \end{equation*} It can be
shown that this formal writing corresponds to the supersymmetric completion of
the Wess-Zumino term that describes the coupling of the D7 brane gauge fields to the Ramond-Ramond
form $C_4$.}, so that altogether we have
\begin{equation}
\label{acsuptree}
S_{\mathrm{tree}}(M,T) = \frac{1}{(2\pi)^4} \int \!d^8x\, d^8\theta\,
F_{\mathrm{tree}}(M,T) + \mathrm{c.c.}~,
\end{equation}
with
\begin{equation}
\label{suptree}
F_{\mathrm{tree}}(M,T)=2\pi\ii \sum_{\ell=1}^\infty 
\frac{(2\pi\alpha')^{2\ell-4} \,\tr M^{2\ell}}{(2\ell)!} \,
\frac{\partial^{2\ell-4}\, T}{\partial z^{2\ell-4}}~.
\end{equation}
The action (\ref{acsuptree}) reduces to the source action
$S_{\mathrm{source}}$ if we single out the terms linear in $\widetilde{\overline\tau}$,
appearing in $T$ accompanied by $\theta^8$, and then set the fields to their
classical values. In other words, the current $J_{\mathrm{cl}}$ of
\eq{sbst} is related to the prepotential by 
\begin{equation}
\label{stcl}
J_{\mathrm{cl}} =-\frac{(2\pi\alpha')^4}{2\pi}\, \frac{\delta
F_{\mathrm{tree}}(M,T)}{\delta\big(\theta^8\,\widetilde{\overline\tau}\big)}
\Bigg|_{T=\tau_0\,,\,M=m_{\mathrm{cl}}}
\equiv -\frac{(2\pi\alpha')^4}{2\pi}\,\bar\delta F_{\mathrm{tree}}(M,T)~,
\end{equation}
where in the second step to simplify our notations we have defined the operation
$\bar\delta$ acting on any object $\star$ as follows:
\begin{equation}
\bar\delta\,\star \equiv \frac{\delta\,
\star}{\delta\big(\theta^8\,\widetilde{\overline\tau}\big)}\Bigg|_{T=\tau_0\,,\,M=m_{\mathrm{cl}}}~.
\label{bardelta}
\end{equation}
Notice that the current $J_{\mathrm{cl}}$ defined by \eq{stcl} is dimensionless
since $\bar\delta F_{\mathrm{tree}}$ has dimensions of (length)$^{-8}$; indeed, if $\star$ has scaling dimensions of (length)$^{\nu}$, then $\bar\delta\,\star$ has scaling dimensions of (length)$^{\nu-4}$.
Using this definition, it is easy to check that by applying the operation $\bar\delta$ to the
prepotential (\ref{suptree}), the current (\ref{Jpis}) is correctly reproduced.

If in Eq.s (\ref{acsuptree}) and (\ref{suptree}) we take all $\theta$'s from the superfield $T$,
we obtain the source terms (\ref{smtau}). These, however, are linked by supersymmetry 
to terms, among others, with the schematic structure
\begin{equation}
\label{Fmtau}
F^4\, m^{2\ell-4} \frac{\partial^{2\ell-4}\, \tau}{\partial z^{2\ell-4}}
\end{equation}
arising when we select all $\theta$'s from the $M^{2\ell}$ part.
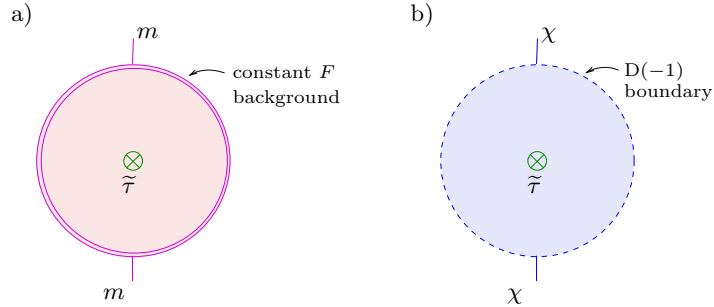
\begin{figure}
\begin{center}
\begin{picture}(0,0)%
\includegraphics{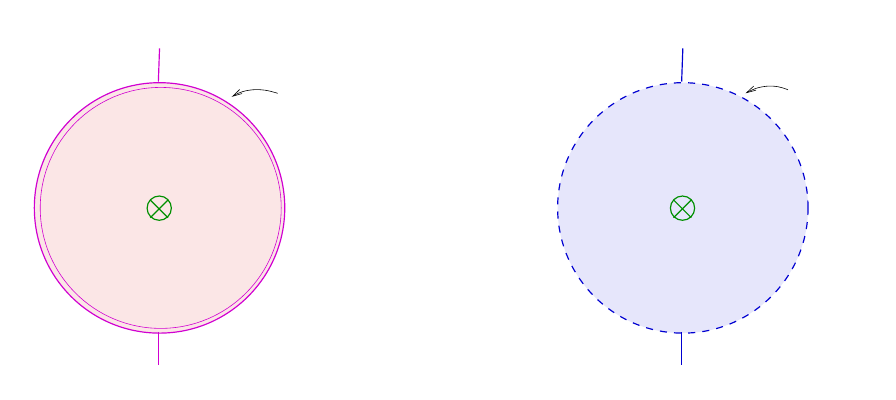}%
\end{picture}%
\setlength{\unitlength}{1492sp}%
\begingroup\makeatletter\ifx\SetFigFont\undefined%
\gdef\SetFigFont#1#2#3#4#5{%
  \reset@font\fontsize{#1}{#2pt}%
  \fontfamily{#3}\fontseries{#4}\fontshape{#5}%
  \selectfont}%
\fi\endgroup%
\begin{picture}(11340,5044)(121,-4307)
\put(8296,-4171){\makebox(0,0)[lb]{\smash{{\SetFigFont{9}{10.8}{\familydefault}{\mddefault}{\updefault}$\chi$}}}}
\put(8851,164){\makebox(0,0)[lb]{\smash{{\SetFigFont{9}{10.8}{\familydefault}{\mddefault}{\updefault}$\chi$}}}}
\put(1651,-4171){\makebox(0,0)[lb]{\smash{{\SetFigFont{9}{10.8}{\familydefault}{\mddefault}{\updefault}$m$}}}}
\put(1956,-2416){\makebox(0,0)[lb]{\smash{{\SetFigFont{9}{10.8}{\familydefault}{\mddefault}{\updefault}${\widetilde\tau}$}}}}
\put(2206,164){\makebox(0,0)[lb]{\smash{{\SetFigFont{9}{10.8}{\familydefault}{\mddefault}{\updefault}$m$}}}}
\put(8626,-2416){\makebox(0,0)[lb]{\smash{{\SetFigFont{9}{10.8}{\familydefault}{\mddefault}{\updefault}${\widetilde\tau}$}}}}
\put(3781,-511){\makebox(0,0)[lb]{\smash{{\SetFigFont{7}{8.4}{\familydefault}{\mddefault}{\updefault}constant $F$}}}}
\put(3781,-916){\makebox(0,0)[lb]{\smash{{\SetFigFont{7}{8.4}{\familydefault}{\mddefault}{\updefault}background}}}}
\put(10216,-466){\makebox(0,0)[lb]{\smash{{\SetFigFont{7}{8.4}{\familydefault}{\mddefault}{\updefault}D$(-1)$}}}}
\put(10216,-826){\makebox(0,0)[lb]{\smash{{\SetFigFont{7}{8.4}{\familydefault}{\mddefault}{\updefault}boundary}}}}
\put(136,434){\makebox(0,0)[lb]{\smash{{\SetFigFont{9}{10.8}{\familydefault}{\mddefault}{\updefault}a)}}}}
\put(6706,434){\makebox(0,0)[lb]{\smash{{\SetFigFont{9}{10.8}{\familydefault}{\mddefault}{\updefault}b)}}}}
\end{picture}%
\end{center}
\caption{In a constant-$F$ background, indicated in the left figure by a double
line for the disk boundary, there is a non-vanishing amplitude with two $m$
scalar fields and the axio-dilaton, proportional to $\bar p^2$ (this is the
$\ell=3$ case of \eq{Fmtau}). In the background represented by a D-instanton, this
diagram has a counterpart, represented by the diagram on the right. In this
latter, the blue dashed line indicates that the boundary of the disk is attached
to D-instantons, and there is the insertions of two $\chi$ moduli, whose vertex
have the same expression of the $m$ vertices. See Section 
\ref{sec:dcmoduli} for details. 
}
\label{fig:chitau}
\end{figure}
The presence of such interaction terms indicates that, should we turn on a
background gauge field $F$ on the D7 branes, we would find a non-vanishing disk
amplitude for the emission of $(2\ell-4)$ open vertices for $m$ and one closed
$\tau$ vertex, proportional to $\bar p^{2\ell-4}$, as represented in Fig.
\ref{fig:chitau}a). Instead of a constant background, we could consider an
instanton solution of the quartic theory, which, as discussed in
\cite{Billo':2009gc}, is represented by a D-instanton. For a disk with its
boundary attached to a D-instanton we expect therefore to find a non-vanishing
amplitude, proportional to $\bar p^{2\ell-4}$, with the insertion of $(2\ell-4)$ moduli
whose vertex formally has the same structure of the vertex of $m$ and of one
closed string vertex for $\tau$, as represented in Fig \ref{fig:chitau}b). As we will see
in the next section, such amplitudes, as well as the analogous ones with higher number
of insertions, are indeed present and play a key r\^ole in the computation of
the non-perturbative corrections to the axio-dilaton profile. 

\paragraph{The quantum-corrected source terms}

Let us summarize the situation. We are interested in the source terms for the
axio-dilaton field $\tau$ induced by the presence of the D7 branes. At tree-level,
the effective action on the D7 branes for the open and closed string massless
sector is given (up to non-linear terms in the closed string fluctuations)
by $S_{\mathrm{tree}}(M,T)$ of \eq{acsuptree}. 

Integrating out the open string sector one obtains the contribution of the
D7 branes to the effective action for the closed string degrees of
freedom:
\begin{equation}
 \label{efftau}
 \ee^{-\Gamma(T)} = \int \!\mathcal{D}M\, \ee^{-S_{\mathrm{tree}}(M,T)}~,
 \end{equation}
where $\mathcal{D}M$ denotes the functional integration measure over all the massless
open degrees of freedom appearing in the superfield $M$. With $\Gamma(T)$ we indicate,
with a slight abuse of notation, the effective action for the closed string degrees of freedom,
even if we do not want to claim that they only appear arranged in the superfield $T$.
The effective action $\Gamma(T)$ in general depends on the classical values of the open
string fields, like for example $m_{\mathrm{cl}}$, and contains source terms linear in 
the closed string fluctuations, and in particular in $\widetilde{\overline\tau}$. 
{From} \eq{efftau} it follows  that such effective source terms
can be expressed as the quantum expectation value of the interaction terms $S_{\mathrm{tree}}(M,\widetilde{\overline\tau})$ 
linear in $\widetilde{\overline\tau}$, given in \eq{smtau}, with respect to the usual 
D7 effective action for the open string modes $M$ at fixed $\tau_0$, namely
\begin{equation}
\label{qefftau}
 \Gamma_{\mathrm{source}}  = \int \!\mathcal{D}M\, S_{\mathrm{tree}}(M,\widetilde{\overline\tau})\, 
 \ee^{-S_{\mathrm{tree}}(M,\tau_0) } + \mathrm{c.c.} ~
 = ~\vev{S_{\mathrm{tree}}(M,\widetilde{\overline\tau})} + \mathrm{c.c.}
\end{equation}
Writing
\begin{equation}
 \label{gbst}
 \Gamma_{\mathrm{source}} = - \frac{T_7}{\tilde\kappa} \int d^8x \, J\,
 \widetilde{\overline\tau} + \mathrm{c.c}
 \end{equation}
in analogy to what we did in \eq{sbst} at the classical level, and using the explicit
expression (\ref{smtau}), we deduce that the quantum current $J$ (in momentum
space) is
\begin{equation}
\label{Jquantumft}
J = - 2 \ii  \sum_{\ell=1}^\infty (-1)^\ell\,
\frac{(2\pi\alpha')^{2\ell}\,\vev{\tr m^{2\ell}}}{(2\ell)!} \,{\bar p}^{2\ell}~.
\end{equation}
The quantum current $J$ is therefore obtained from the classical one given in \eq{Jft} 
by promoting the classical values $m_{\mathrm{cl}}^{2\ell}$ of the field $m$ to the corresponding
quantum expectation values $\vev{\tr m^{2\ell}}$, which are the elements of the
so-called chiral ring of the D7 brane theory. 

Another way to express the quantum current is by writing the effective action
$\Gamma$ in terms of a superpotential:
\begin{equation}
\label{effsup}
\Gamma = \frac{1}{(2\pi)^4} \int d^8x\, d^8\theta\,
F + \mathrm{c.c.}
\end{equation}
where $F$ can have classical, perturbative and non-perturbative parts:
\begin{equation}
\label{fcpnp}
F = F_{\mathrm{tree}} + F_{\mathrm{pert}} + F_{\mathrm{n.p.}}~.
\end{equation}
The classical term $F_{\mathrm{tree}}$ is just the one given in \eq{suptree}
with $M$ frozen to $m_{\mathrm{cl}}$. Due to the high degree
of supersymmetry, the perturbative contributions to the superpotential vanish.
Non-perturbative terms correspond to the contributions of topologically
non-trivial sectors in the functional integral (\ref{efftau}) and are
represented in the string picture
by the inclusion of  D-instantons in the background.

The source terms $\Gamma_{\mathrm{source}}$ are obtained by selecting in $\Gamma$ the
terms linear in $\widetilde{\overline\tau}$, and then setting the closed string
fields to their classical value, {\it{i.e.}} $T=\tau_0$. If we start from the superpotential, we must
also single out the terms proportional to $\theta^8$. Therefore to get $\Gamma_{\mathrm{source}}$
only the combination $\theta^8\widetilde{\overline\tau}$ is relevant. This is precisely the 
combination appearing inside the superfield $T$, and thus we can ignore all other possible
dependencies on $\widetilde{\overline\tau}$ that may arise in the effective action.
Therefore we can write
\begin{equation}
\label{stquantum}
J = -\frac{(2\pi\alpha')^4}{2\pi}\,\bar\delta F~,
\end{equation}
where $\bar\delta$ is the variation operator defined in \eq{bardelta}.

In the next sections we will compute the non-perturbative prepotential
$F_{\mathrm{n.p.}}$ induced by the presence of D-instantons, including the
couplings to the closed string fields, and will use \eq{stquantum} to derive the
corresponding source terms for the axio-dilaton, from which the non-perturbative
corrections to its profile follow. In doing so, we expect to find that the
effective source is obtained by taking into account the instanton corrections to
the chiral ring elements, thus verifying the heuristic discussion that led us to
\eq{Jquantumft}.

\section{Axio-dilaton couplings in the D-instanton action}
\label{sec:dcmoduli}
The non-perturbative sectors of the SO(8) 
field theory living on the D7 branes can be described
by adding $k$ D-instantons in the same orientifold fixed point occupied by the D7's
\cite{Billo':2009gc,Billo:2009di}.
These D-instantons are sources for the Ramond-Ramond scalar $\widetilde C_0$.
Thus, from the Wess-Zumino part of the D7 action (\ref{quarticgf}), we see that $k$
D(--1) branes correspond to a gauge field configuration with fourth Chern number $k$.
Moreover, for this configuration the classical quartic action reduces to $k$ 
times the D-instanton action.

The physical excitations of the open strings with at least one end-point on the
D(--1) branes account for the moduli of such instanton configurations which will be
collectively denoted $\cM_{(k)}$. 
They comprise the neutral sector, corresponding to D(--1)/D(--1) open strings,
which contains those moduli that do not transform under the SO(8) gauge group, 
and the charged sector, arising from D(--1)/D7 open strings, which includes those moduli
that transform in the fundamental representation of SO(8).

The neutral moduli are a vector $a_\mu$ and a complex scalar $\chi$
(plus its conjugate $\bar\chi$) in the Neveu-Schwarz sector, and a chiral fermion $\eta_\alpha$
and an anti-chiral fermion $\lambda_{\dot\alpha}$ in the Ramond sector.
The bosonic moduli have canonical dimensions of (length)$^{-1}$, while the
fermionic ones have canonical dimensions of (length)$^{-\frac{3}{2}}$.
All these neutral moduli are $k\times k$ matrices, but the consistency with the
orientifold projection on the D7 branes requires that $\chi$, $\bar\chi$ and
$\lambda_{\dot\alpha}$ transform in the anti-symmetric (or adjoint)
representation of $\mathrm{SO}(k)$, while $a_\mu$ and $\eta_\alpha$ are in the
symmetric one. The diagonal parts of $a_\mu$ and $\eta_\alpha$ represent the
bosonic and fermionic Goldstone modes of the (super)translations of the
D7 branes world-volume that are broken by the D-instantons and thus can be
identified, respectively, with the bosonic and fermionic coordinates $x_\mu$ and
$\theta_\alpha$ of the $8d$ superspace. More explicitly, we have
\begin{equation}
 x_\mu \sim \ell_{s}^2\, \mathrm{tr}(a_\mu)\quad,\quad
\theta_\alpha \sim \ell_{s}^2\, \mathrm{tr}(\eta_\alpha)~,
\label{xtheta}
\end{equation}
where the factors of the string length $\ell_{s}=\sqrt{\alpha'}$ 
have been introduced to give $x_\mu$ and $\theta_\alpha$ their 
standard dimensions.

Since the D(--1)/D7 open strings have eight directions with mixed Dirichlet-Neumann boundary 
conditions, there are no bosonic charged excitations that satisfy the physical state condition in
the Neveu-Schwarz sector.
On the other hand, the fermionic Ramond sector of the D(--1)/D7 system contains
physical moduli, denoted as $\mu$ and $\bar\mu$ depending on the orientation.
They are, respectively, $k\times N$ and $N\times k$ matrices (with $N=8$ in our
specific case). Since the orientifold parity (\ref{typeipd}) exchanges
the two orientations, $\mu$ and $\bar\mu$ 
are related according to $\bar\mu = -
\,{}^{\mathrm{t}}\!\mu$.
For all the physical moduli listed above,
it is possible to write vertex operators of conformal dimension 1 that are collected in Table~1.
\begin{table}[htb]
\label{tab:10}
\begin{center}
 \begin{tabular}{|c|c|c|c|c|
}
\hline
 {\phantom{\vdots}}&$\mathrm{SO}(k)$&
 $\mathrm{SO}(8)$ & 
dimensions & vertex\\
\hline
${\phantom{\vdots}}a_\mu$&$\mathrm{symm}$ & $\mathbf{1}$ & 
(length)$^{-1}$ &
$V_a =  \ell_s \,a_\mu \, \psi^\mu\,\ee^{-\varphi}~~~~~~~~~$ \\
${\phantom{\vdots}}\eta_\alpha$ & $\mathrm{symm}$ & $\mathbf{1}$ & 
(length)$^{-3/2}$ & 
$V_\eta = \ell_s^{3/2}\, \eta_{\alpha}\,  S^{\alpha -} \,\ee^{-\frac{1}{2}\varphi}~~$\\
${\phantom{\vdots}}\lambda_{\dot \alpha}$ & $\mathrm{adj}$ & $\mathbf{1}$ & 
(length)$^{-3/2}$ & 
$V_{\lambda} = \ell_s^{3/2}\, \lambda_{\dot \alpha}\, S^{\dot \alpha +}\,
\ee^{-\frac{1}{2}\varphi}~~$ \\
${\phantom{\vdots}}\bar\chi$&$\mathrm{adj}$ & $\mathbf{1}$ & 
(length)$^{-1}$ & 
$V_{\bar \chi} = \ell_s\,{\bar \chi}\,  \Psi\,\ee^{-\varphi}~~~~~~~~~~~~$\\
${\phantom{\vdots}}\chi$&$\mathrm{adj}$ & $\mathbf{1}$ & 
(length)$^{-1}$ & 
$V_\chi = \ell_s\, \chi \, {\overline \Psi}\,\ee^{-\varphi}~~~~~~~~~~~~$\\
${\phantom{\vdots}}\mu$&$\mathbf{k}$ & $\mathbf{8}_v$ & 
(length)$^{-3/2}$ & 
$V_\mu = \ell_s^{3/2}\, \mu\,  \Delta\,S^+ \,\ee^{-\frac{1}{2}\varphi}~~~~$\\
\hline
\end{tabular}
\caption{Transformation properties, scaling dimensions and vertices of the moduli in the 
D(--1)/D7 system. In the last column,
$\varphi$ is the bosonic field appearing in the bosonization of the superghost, the
$S$'s are spin fields with chiral (anti-chiral) indices $\alpha (\dot\alpha)$ and $+ (-)$ in
the first eight and last two directions respectively, while $\Delta$ is the bosonic twist
field in the first eight directions. The factors of the string length $\ell_s=\sqrt{\alpha'}$
are inserted in order to make the vertex operators dimensionless.
}
\label{tab:mod}
\end{center}
\end{table}

The D-instanton action can be derived by computing correlation functions
on disks with at least part of their boundary lying on the D(--1) branes; 
for details in the type I$^\prime$ model
we refer to Ref.s~\cite{Billo':2009gc,Billo:2009di} and here we only recall
the results that are most significant for our present purposes.

The first contribution to the moduli action is the classical part
\begin{equation}
S_{\mathrm{cl}}= \frac{2\pi k}{g_s} =- 2 \pi \ii \,k \,\tau_0
\label{S01}
\end{equation}
corresponding to the (topological) normalization of disk amplitudes
with D(--1) boundary conditions.
This term accounts for disk amplitudes on $k$ D-instantons with no moduli
insertions \cite{Polchinski:1994fq,Billo:2002hm}.

The second type of contributions originates from disks amplitudes
involving the vertex operators of the D(--1)/D(--1) or D(--1)/D7 strings.
They lead to the following moduli action \cite{Billo':2009gc,Billo:2009di}
\begin{equation}
\begin{aligned}
S_{\mathrm{mod}}(\cM_{(k)})
= &\,\frac{1}{g_0^2}\,
\mathrm{tr}\Bigg\{ \ii\,\lambda_{\dot\alpha}
\gamma_\mu^{\dot\alpha\beta} [a^\mu,\eta_\beta] -
\frac{\ii}{\sqrt 2}\,\lambda_{\dot\alpha}[\chi,\lambda^{\dot\alpha}] 
- \frac{\ii}{\sqrt 2}\,\eta^\alpha[\bar\chi,\eta_\alpha]
\,-\ii \,{\sqrt 2}\, {}^{\mathrm{t}}\!\mu \,\chi\,\mu \\ &~~~~~~~~
-\frac{1}{4}[a_\mu,a_\nu][a^\mu,a^\nu] - [a_\mu,\chi][a^\mu,\bar\chi]
-\frac{1}{2}[\chi,\bar\chi]^2
\Bigg\}
\end{aligned}
\label{S1}
\end{equation}
where $g_0^2=g_s\,(4\pi^3\alpha'^2)^{-1}$ 
is the Yang-Mills coupling constant of the zero-dimensional gauge theory
defined on the D-instantons. Notice that this action does not depend on the diagonal components
$x_\mu$ and $\theta_\alpha$, defined in \eq{xtheta}, which are true zero-modes of the instanton
system. 
The quartic interactions $[a_\mu,a_\nu][a^\mu,a^\nu]$
can be disentangled by introducing seven auxiliary fields $D_m$ ($m=1,\cdots,7$) with
canonical dimension of (length)$^{-2}$,
and replacing $S_{\mathrm{mod}}(\cM_{(k)})$ with
\begin{equation}
\begin{aligned}
S'_{\mathrm{mod}}(\cM_{(k)})
= &\,\frac{1}{g_0^2}\,
\mathrm{tr}\Bigg\{ \ii\,\lambda_{\dot\alpha}
\gamma_\mu^{\dot\alpha\beta} [a^\mu,\eta_\beta] -
\frac{\ii}{\sqrt 2}\,\lambda_{\dot\alpha}[\chi,\lambda^{\dot\alpha}] 
- \frac{\ii}{\sqrt 2}\,\eta^\alpha[\bar\chi,\eta_\alpha] 
\,-\ii \,{\sqrt 2}\, {}^{\mathrm{t}}\!\mu \,\chi\,\mu \\ &~~~~~~~
+\frac{1}{2}D_m D^m-\frac{1}{2}D_m(\tau^m)_{\mu\nu}\,[a^\mu,a^\nu] 
- [a_\mu,\chi][a^\mu,\bar\chi]
-\frac{1}{2}[\chi,\bar\chi]^2
\Bigg\}~.
\end{aligned}
\label{S1a}
\end{equation}
Here $(\tau^m)_{\mu\nu}$ are the seven $\gamma$-matrices of SO(7), implying that the eight-dimensional
vector indices $\mu,\nu,\cdots$ of SO(8) can be interpreted also as spinorial indices of SO(7).
By eliminating the auxiliary fields $D_m$ through their equations of motion and exploiting the
properties of the $\tau^m$ matrices, one can easily check that $S'_{\mathrm{mod}}(\cM_{(k)})$ and
$S_{\mathrm{mod}}(\cM_{(k)})$ are equivalent. {From} now on, we will work with the action (\ref{S1a}) and the enlarged set of moduli that includes also the auxiliary fields $D_m$.

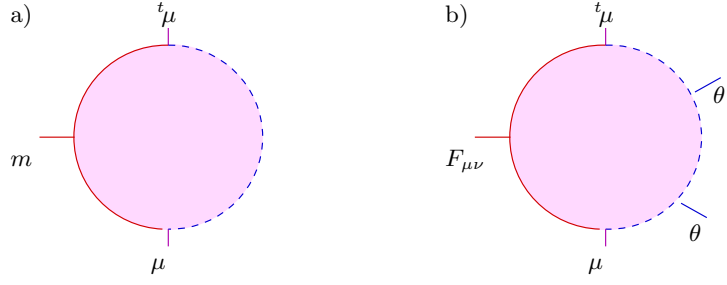
\begin{figure}
\begin{center}
\begin{picture}(0,0)%
\includegraphics{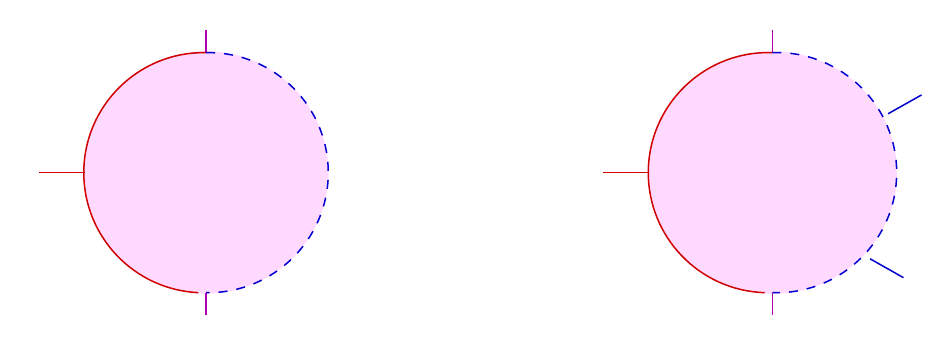}%
\end{picture}%
\setlength{\unitlength}{1895sp}%
\begingroup\makeatletter\ifx\SetFigFontNFSS\undefined%
\gdef\SetFigFontNFSS#1#2#3#4#5{%
  \reset@font\fontsize{#1}{#2pt}%
  \fontfamily{#3}\fontseries{#4}\fontshape{#5}%
  \selectfont}%
\fi\endgroup%
\begin{picture}(9239,3539)(-275,-3000)
\put(1560,-2911){\makebox(0,0)[lb]{\smash{{\SetFigFontNFSS{9}{10.8}{\familydefault}{\mddefault}{\updefault}$\mu$}}}}
\put(1635,314){\makebox(0,0)[lb]{\smash{{\SetFigFontNFSS{9}{10.8}{\familydefault}{\mddefault}{\updefault}${}^t\!\mu$}}}}
\put(-260,-1561){\makebox(0,0)[lb]{\smash{{\SetFigFontNFSS{9}{10.8}{\familydefault}{\mddefault}{\updefault}$m$}}}}
\put(-260,314){\makebox(0,0)[lb]{\smash{{\SetFigFontNFSS{9}{10.8}{\familydefault}{\mddefault}{\updefault}a)}}}}
\put(7297,314){\makebox(0,0)[lb]{\smash{{\SetFigFontNFSS{9}{10.8}{\familydefault}{\mddefault}{\updefault}${}^t\!\mu$}}}}
\put(5371,-1561){\makebox(0,0)[lb]{\smash{{\SetFigFontNFSS{9}{10.8}{\familydefault}{\mddefault}{\updefault}$F_{\mu\nu}$}}}}
\put(5371,314){\makebox(0,0)[lb]{\smash{{\SetFigFontNFSS{9}{10.8}{\familydefault}{\mddefault}{\updefault}b)}}}}
\put(7220,-2911){\makebox(0,0)[lb]{\smash{{\SetFigFontNFSS{9}{10.8}{\familydefault}{\mddefault}{\updefault}$\mu$}}}}
\put(8530,-2536){\makebox(0,0)[lb]{\smash{{\SetFigFontNFSS{9}{10.8}{\familydefault}{\mddefault}{\updefault}$\theta$}}}}
\put(8838,-736){\makebox(0,0)[lb]{\smash{{\SetFigFontNFSS{9}{10.8}{\familydefault}{\mddefault}{\updefault}$\theta$}}}}
\end{picture}%
\end{center}
\caption{a) The disk diagram describing the interaction of the D-instantons with the
adjoint scalar field $m$ living on the D7 branes. b) The susy broken by the instanton relates
the previous diagram to a diagram involving the field-strength $F_{\mu\nu}$ with two extra insertions of the vertices for the modulus $\theta$, effectively promoting $m(x)$
to the superfield $M(x,\theta)$.}
\label{fig:mumum}
\end{figure} 
A further type of contributions to the instanton action comes from the
interactions among the instanton moduli and the gauge fields
propagating on the world-volume of the D7 branes, which we have combined into 
the super-field (\ref{defM}). Such interactions can be obtained by computing mixed disk amplitudes
involving the vertex operators for the charged moduli and the vertex operators for the
dynamical fields. For example, in Fig.~\ref{fig:mumum}a) we have represented 
the mixed D(--1)/D7 diagram describing the coupling between the complex scalar $m$ with two charged 
moduli $\mu$, and yielding a term in the action proportional to $\mathrm{tr}\left\{{}^{\mathrm{t}}\!\mu\,\mu\,m\right\}$. 
By exploiting the identification (\ref{xtheta}) between the fermionic superspace coordinates
$\theta_\alpha$ and the trace part of $\eta_\alpha$, it is immediate
to realize that the couplings with the higher components of the superfield $M(x,\theta)$ 
can be obtained from disk diagrams with extra fermionic insertions, like for
instance the one involving the gauge field strength $F_{\mu\nu}$ and two $\theta$'s 
represented in Fig.~\ref{fig:mumum}b). Equivalently, these interactions can be obtained by
acting with the broken supersymmetry transformations on the lower terms; all in all this
amounts to replace the moduli action (\ref{S1a}) with
\begin{equation}
S'_{\mathrm{mod}}(\cM_{(k)}, M)
= S'_{\mathrm{mod}}(\cM_{(k)})\,
+\,\frac{\ii\sqrt 2}{g_0^2}
\mathrm{tr}\Big\{{}^{\mathrm{t}}\!\mu\,\mu\,M(x,\theta)\Big\}~.
\label{smod12}
\end{equation}

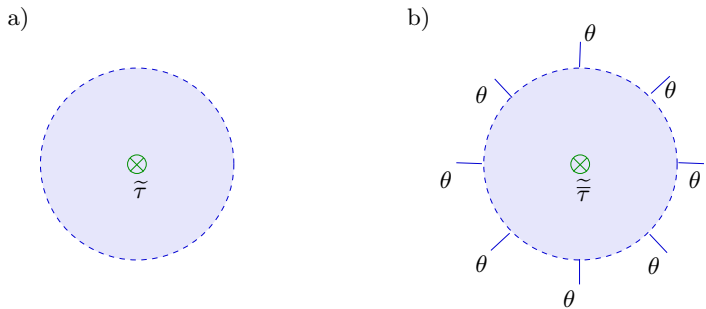
\begin{figure}
\begin{center}
\begin{picture}(0,0)%
\includegraphics{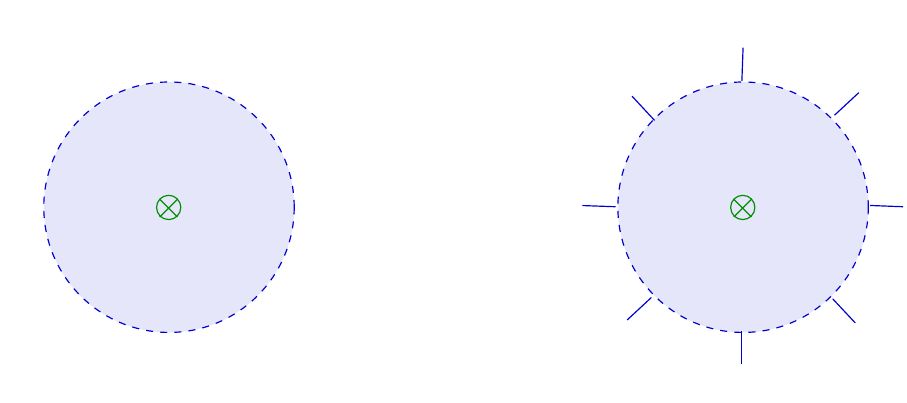}%
\end{picture}%
\setlength{\unitlength}{1492sp}%
\begingroup\makeatletter\ifx\SetFigFont\undefined%
\gdef\SetFigFont#1#2#3#4#5{%
  \reset@font\fontsize{#1}{#2pt}%
  \fontfamily{#3}\fontseries{#4}\fontshape{#5}%
  \selectfont}%
\fi\endgroup%
\begin{picture}(11489,5085)(121,-4360)
\put(136,434){\makebox(0,0)[lb]{\smash{{\SetFigFont{9}{10.8}{\familydefault}{\mddefault}{\updefault}a)}}}}
\put(6706,434){\makebox(0,0)[lb]{\smash{{\SetFigFont{9}{10.8}{\familydefault}{\mddefault}{\updefault}b)}}}}
\put(2206,-2446){\makebox(0,0)[lb]{\smash{{\SetFigFont{9}{10.8}{\familydefault}{\mddefault}{\updefault}$\widetilde\tau$}}}}
\put(7831,-871){\makebox(0,0)[lb]{\smash{{\SetFigFont{9}{10.8}{\familydefault}{\mddefault}{\updefault}$\theta$}}}}
\put(9616,164){\makebox(0,0)[lb]{\smash{{\SetFigFont{9}{10.8}{\familydefault}{\mddefault}{\updefault}$\theta$}}}}
\put(7246,-2266){\makebox(0,0)[lb]{\smash{{\SetFigFont{9}{10.8}{\familydefault}{\mddefault}{\updefault}$\theta$}}}}
\put(9496,-2496){\makebox(0,0)[lb]{\smash{{\SetFigFont{9}{10.8}{\familydefault}{\mddefault}{\updefault}$\widetilde{\overline\tau}$}}}}
\put(10936,-781){\makebox(0,0)[lb]{\smash{{\SetFigFont{9}{10.8}{\familydefault}{\mddefault}{\updefault}$\theta$}}}}
\put(11341,-2266){\makebox(0,0)[lb]{\smash{{\SetFigFont{9}{10.8}{\familydefault}{\mddefault}{\updefault}$\theta$}}}}
\put(10666,-3706){\makebox(0,0)[lb]{\smash{{\SetFigFont{9}{10.8}{\familydefault}{\mddefault}{\updefault}$\theta$}}}}
\put(9271,-4246){\makebox(0,0)[lb]{\smash{{\SetFigFont{9}{10.8}{\familydefault}{\mddefault}{\updefault}$\theta$}}}}
\put(7831,-3661){\makebox(0,0)[lb]{\smash{{\SetFigFont{9}{10.8}{\familydefault}{\mddefault}{\updefault}$\theta$}}}}
\end{picture}%
\end{center}
\caption{a) The disk diagram describing the interaction of the D-instantons with the axio-dilaton vertex. b) The diagram, linked by supersymmetry to the previous one, which
comprises the insertion of eight $\theta$-vertices and a vertex for $\bar\tau$.}
\label{fig:tautheta}
\end{figure}

The last type of contributions to the instanton action comes from the interactions among the
moduli and the closed string excitations describing the gravitational degrees of freedom
propagating in the bulk. The simplest of such contributions corresponds to the coupling
of the axio-dilaton fluctuations to the D-instantons which can be obtained, for example,
by considering the tree-level Born-Infeld and Wess-Zumino actions of $k$ D-instantons, or
equivalently by computing the overlap between the D(--1) boundary state and the vertex operator
of the axio-dilaton field represented by the disk diagram of Fig.~\ref{fig:tautheta}a).
This amounts to simply replace the classical instanton action (\ref{S01}) with
\begin{equation}
S_{\mathrm{cl}}= - 2 \pi \ii \,k \,\tau
\label{S01a}
\end{equation}
{\it i.e.} to promote the expectation value $\tau_0$ to the entire axio-dilaton field $\tau$.
Actually, by exploiting the broken supersymmetries we can do more and further promote $\tau$ to
the full-fledged superfield $T$ introduced in \eq{Tsuper}. Thus, if we include all couplings
among the D-instantons and the closed string degrees of freedom, the classical action
(\ref{S01a}) becomes
\begin{equation}
S'_{\mathrm{cl}}= - 2 \pi \ii \,k \,T = - 2 \pi \ii \,k \,\tau_0 - 2 \pi \ii \,k \,\widetilde T~.
\label{S01b}
\end{equation}
Notice that, among others, this action contains an interaction term with the following structure
\begin{equation}
 \theta^8\,\bar p^4 \,\widetilde{\overline \tau}
\label{theta8bartau}
\end{equation}
which accounts for the coupling among eight superspace coordinates 
and the (complex conjugate) axio-dilaton.
This term is represented by the diagram in Fig.~\ref{fig:tautheta}b) and
can be regarded as the highest supersymmetric partner of the simple emission diagram
in Fig.~\ref{fig:tautheta}a). In particular the coupling (\ref{theta8bartau}) implies the existence
of a non-trivial amplitude involving eight vertices for the $\theta$'s and one vertex for the
axion field $\widetilde C_0$. In Appendix \ref{app:sd} we compute this amplitude using
conformal field theory techniques, thus finding an explicit confirmation of the above structure.

As argued in Section \ref{treecd7}, in the presence of a non-constant axio-dilaton profile
we expect further interaction terms in the moduli action involving the derivatives of $\tau$.
In particular, in the present configuration we expect to find couplings with the following schematic
structure
\begin{equation}
 \tr (\chi^{2\ell})\,\frac{\partial^{2\ell}\widetilde \tau}{\partial z^{2\ell}} ~~~~\mbox{or}~~~~ 
\tr (\chi^{2\ell})\,\bar p^{2\ell}\,\widetilde \tau~,
\label{chitau1}
\end{equation}
which are the D-instanton analogues of the interactions (\ref{Fmtau})
that are present on the D7 branes (see also Fig.~\ref{fig:chitau}). 
To check this idea we can compute 
the mixed open/closed disk amplitude with $2\ell$ vertices for $\chi$ and one vertex for the axion
$\widetilde C_0$. Since we are interested in effects depending on the $z$-derivatives of $\tau$, it is
easier to work with the vertex for the emission of the Ramond-Ramond field strength
$F_z= \partial_z\,\widetilde C_0 \equiv \ii\,\bar p\,\widetilde C_0$, given by
\begin{equation}
\label{rr}
V_{\widetilde C_0}(w,\bar w) \,=\, \frac{2\pi g_s\ell_s}{8}\,\ii\,\bar p\,\widetilde C_0\,
S^{\dot\alpha +}(w)\, \ee^{\ii\,\ell_s\bar p  Z(w)}\,\ee^{-\frac{1}{2}\varphi(w)}~
\widetilde{S}_{\dot\alpha}^{\phantom{\dot\alpha}\!+}(\bar w) \, \ee^{\ii\,\ell_s\bar p \widetilde{Z}(\bar w)} 
\,\ee^{-\frac{1}{2}\widetilde{\varphi}(\bar w)}~.
\end{equation}
Here we have included the normalization prefactor $(2\pi g_s\ell_s)/8$ in order to be consistent
with the bulk action (\ref{bulk}) and used the
charge conjugation matrix $C_{\dot\alpha\dot\beta}$
to contract the antichiral spinor indices of the left and right spin
fields $S$ and $\widetilde S$ (see Appendix \ref{app:notconv} for some details). 
Moreover, we have assumed that the axion depends only on $z$ and thus
have used only the momentum $\bar p$ in its emission vertex. 
This is clearly a simplification 
since in general we could allow also for a dependence on $\bar z$ (and hence also 
on $p$); however, this is enough for our purposes since, as discussed in
detail in Section \ref{sec:dcd7}, we are after the
terms with holomorphic derivatives of the axio-dilaton field%
\footnote{Notice that we do not consider either any dependence of the closed string background
on the momenta $p_\mu$ along the D7 brane world-volume in order not to generate interactions
in the instanton action that would lead to terms breaking the eight-dimensional Lorentz invariance.}.

A quick inspection shows that the only non-trivial couplings are among $\widetilde C_0$ and
an even number of $\chi$ moduli due to the Chan-Paton structure of the latter.
Usually, the calculation of amplitudes involving a large number of vertex operators
is a daunting task, but in this particular case it can be carried out explicitly for an 
arbitrary number of $\chi$ insertions: indeed, in order to soak up the superghost number anomaly,
all but one vertices $V_\chi$ must be chosen in the 0-superghost picture where they are
extremely simple. The details of these calculations are provided in Appendix \ref{subapp:C0chi},
while here we simply quote the final result which, for the amplitude corresponding to the diagram
in Fig.~\ref{fig:tauchi}, is 
\begin{figure}
\begin{center}
\begin{picture}(0,0)%
\includegraphics{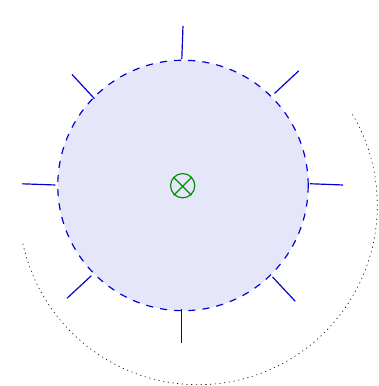}%
\end{picture}%
\setlength{\unitlength}{1492sp}%
\begingroup\makeatletter\ifx\SetFigFont\undefined%
\gdef\SetFigFont#1#2#3#4#5{%
  \reset@font\fontsize{#1}{#2pt}%
  \fontfamily{#3}\fontseries{#4}\fontshape{#5}%
  \selectfont}%
\fi\endgroup%
\begin{picture}(4801,4857)(661,-4279)
\put(1261,-736){\makebox(0,0)[lb]{\smash{{\SetFigFont{9}{10.8}{\familydefault}{\mddefault}{\updefault}$\chi$}}}}
\put(3046,299){\makebox(0,0)[lb]{\smash{{\SetFigFont{9}{10.8}{\familydefault}{\mddefault}{\updefault}$\chi$}}}}
\put(676,-2131){\makebox(0,0)[lb]{\smash{{\SetFigFont{9}{10.8}{\familydefault}{\mddefault}{\updefault}$\chi$}}}}
\put(2926,-2411){\makebox(0,0)[lb]{\smash{{\SetFigFont{9}{10.8}{\familydefault}{\mddefault}{\updefault}$\widetilde C_0$}}}}
\put(4366,-646){\makebox(0,0)[lb]{\smash{{\SetFigFont{9}{10.8}{\familydefault}{\mddefault}{\updefault}$\chi$}}}}
\end{picture}%
\end{center}
\caption{The disk diagram describing the interaction between any even number of
$\chi$ moduli and the axion vertex.}
\label{fig:tauchi}
\end{figure}
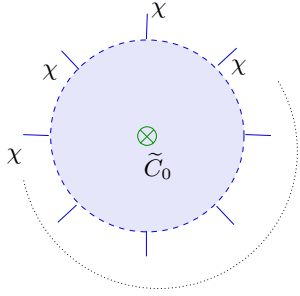  
\begin{equation}
\label{chiC0}
\lvev \underbrace{V_{\chi}\cdots V_\chi}_{2\ell}\, V_{\widetilde C_0} \rvev
=2\pi\ii \,\frac{(2\pi\alpha')^{2\ell}}{(2\ell)!}\,\tr(\chi^{2\ell})\,\bar p^{2\ell}\,\widetilde C_0
~.
\end{equation}
After including the contribution from the dilaton fluctuations, we can conclude that the 
instanton action contains also the following terms%
\footnote{Recall that, with Euclidean signature, scattering amplitudes and 
couplings in the action differ by a minus sign.}
\begin{equation}
\label{stau}
-2\pi\ii\sum_{\ell=1}^{\infty}
\frac{(2\pi\alpha')^{2\ell}}{(2\ell)!}\,\tr(\chi^{2\ell})\,\bar p^{2\ell}\,\widetilde \tau
\end{equation}
in agreement with our expectations. By exploiting the broken supersymmetries or, equivalently,
by inserting vertices for the superspace coordinates $\theta_\alpha$, we can
promote the axion to the complete superfield (\ref{Tsuper}), obtaining 
\begin{equation}
\label{sT}
-2\pi\ii\sum_{\ell=1}^{\infty}
\frac{(2\pi\alpha')^{2\ell}}{(2\ell)!}\,\tr(\chi^{2\ell})\,\bar p^{2\ell}\,\widetilde T~.
\end{equation}
This implies that the instanton moduli action contains, among others, the following terms
\begin{equation}
\label{sT1}
-4\pi\ii\sum_{\ell=1}^{\infty}
\frac{(2\pi\alpha')^{2\ell}}{(2\ell)!}\,\tr(\chi^{2\ell})\,\bar p^{2\ell+4}
\,\theta^8\,\widetilde{\overline \tau}~,
\end{equation}
which, as we will discuss in the next section, are responsible for the non-perturbative corrections
in the $\tau$ profile.

Collecting all contributions we have found, we conclude that the action for $k$ 
D-instantons of type I$^\prime$ in the presence of a non-trivial holomorphic 
axio-dilaton background is given by
\begin{equation}
 S_{\mathrm{inst}} =- 2 \pi \ii \,k \,\tau_0 + \widetilde S_{\mathrm{inst}}
\label{sinst}
\end{equation}
with
\begin{equation}
\widetilde S_{\mathrm{inst}}= S'_{\mathrm{mod}}(\cM_{(k)}, M)
-2\pi\ii\sum_{\ell=0}^{\infty}
\frac{(2\pi\alpha')^{2\ell}}{(2\ell)!}\,\tr(\chi^{2\ell})\,\bar p^{2\ell}\,\widetilde T~.
\label{sinst1}
\end{equation}
Notice that for $\ell=0$ we have $\tr(\chi^{2\ell})=k$; thus the $\ell=0$
term in the above expression is just $-2 \pi \ii \,k \,\widetilde T$, which
is precisely the fluctuation part of the classical instanton action as indicated in \eq{S01b}.

In order to perform the integral over the instanton moduli space and obtain explicit results
at instanton number $k$, we have to exploit the localization formulas and adopt Nekrasov's approach
to the multi-instanton calculus \cite{Nekrasov:2002qd,Nekrasov:2003rj}
(see also Ref.s~\cite{Moore:1998et}\nocite{Bruzzo:2002xf} - \cite{Flume:2002az}).
In our stringy context this corresponds to use a deformed instanton action that is obtained 
by coupling the instanton moduli to a non-trivial
``graviphoton'' background in space-time \cite{Billo:2006jm,Ito:2010vx}.
Since the case we are considering has already been discussed in great
detail in Ref.~\cite{Billo:2009di}, we do not repeat the analysis of the deformation effects
here and simply recall that the graviphoton field-strength
$\mathcal F_{\mu\nu}$ can be aligned along the SO(8) Cartan generators and taken to be of the
following form
\begin{equation}
\label{Ff1}
\mathcal F_{\mu\nu} =-2\ii \begin{pmatrix} 
      \varepsilon_1\sigma_2& 0 &0&0  \cr
     0&\varepsilon_2\sigma_2&0&0 \cr
      0&0&\varepsilon_3\sigma_2& 0\cr
      0&0&0&\varepsilon_4\sigma_2 
     \end{pmatrix} ~~~~\mbox{where}~~\sigma_2=\begin{pmatrix} 
      0& -\ii \cr
     \ii&0     \end{pmatrix}
\end{equation} 
with $\varepsilon_1+\varepsilon_2+\varepsilon_3+\varepsilon_4=0$. This closed string
background modifies the instanton action (\ref{sinst1}) inducing new couplings that can
be explicitly computed from string amplitudes.
In particular the term $S'_{\mathrm{mod}}$ acquires a 
dependence on the deformation parameters $\varepsilon_i$'s; however, the explicit expression
of this modified moduli action is not
necessary in the following, and thus we refer the interested reader to Ref.~\cite{Billo:2009di},
and specifically to Sections 4 and 5 of that paper, where all details can be found.
In the next section we will use this approach to show how D-instantons produce
the non-perturbative corrections to the axio-dilaton profile.

\section{The non-perturbative axio-dilaton profile}
\label{sec:profile}
The effects produced by D-instantons are encoded in the non-perturbative low-energy
effective action of the open and closed string degrees of freedom supported by the D7 branes.
To obtain this action we first introduce the $k$-instanton canonical partition 
function $\widetilde Z_k$, which is given by the following integral over the moduli space
\begin{equation}
 \widetilde Z_k = \int d\cM_{(k)} \,\ee^{-\widetilde S_{\mathrm{inst}}}~.
\label{tildeZk}
\end{equation}
Here $\widetilde S_{\mathrm{inst}}$ is the instanton action (\ref{sinst1}) suitably deformed
by the graviphoton background (\ref{Ff1}), {\it i.e.} with $S'_{\mathrm{mod}}(\cM_{(k)}, M)$
replaced by $S'_{\mathrm{mod}}(\cM_{(k)}, M;\varepsilon_i)$. When the fluctuations of the
open and closed string fields are switched off, $\widetilde Z_k$ reduces to the
ordinary D-instanton partition function $Z_k$; indeed
\begin{equation}
\widetilde Z_k \Big|_{T=\tau_0\,,\,M=m_{\mathrm{cl}}}
~= \int \!d\cM_{(k)} \,\ee^{-S'_{\mathrm{mod}}(\cM_{(k)}, m_{\mathrm{cl}};\varepsilon_i)}
~\equiv\,Z_k~.
\label{zk00}
\end{equation}
The above integral over moduli space can be explicitly computed via localization methods 
using Nekrasov's prescription as done in Ref.s~\cite{Billo:2009di,Fucito:2009rs} for our
brane system.

Summing over all instanton numbers, we obtain the grand-canonical 
instanton partition function 
\begin{equation}
\label{tildeZ}
 \widetilde{\mathcal{Z}}= \sum_{k=0}^\infty q^k\,\widetilde Z_k 
\end{equation}
where we have set $\widetilde Z_0=1$ and $q=\ee^{2\pi\ii\tau_0}$.
Then, following Nekrasov \cite{Nekrasov:2002qd,Nekrasov:2003rj}, 
we define the non-perturbative prepotential induced by the D-instantons on the D7 branes
as follows
\begin{equation}
 \widetilde F_{\mathrm{n.p.}} 
= \lim_{\mathcal{E}\to 0}\, \mathcal{E} \log \widetilde{\mathcal{Z}}
\label{tildeF}
\end{equation}
where $\cE=\varepsilon_1\varepsilon_2\varepsilon_3\varepsilon_4$. Expanding in powers of
$q$ leads to
\begin{equation}
 \widetilde F_{\mathrm{n.p.}} \,=\, \sum_{k=1}^\infty
 q^k\,\widetilde F_k \,=\, \sum_{k=1}^\infty
 q^k\!\int \!d\widehat\cM_{(k)} \,\ee^{-\widetilde S_{\mathrm{inst}}}
\label{tildeFk}
\end{equation}
where in the last term we have exhibited the fact that $\widetilde F_k$ is really an integral over the centered moduli $\widehat\cM_{(k)}$, {\it i.e.} all moduli but the ``center of mass'' coordinates $x_\mu$ and their superpartners $\theta_\alpha$ defined in (\ref{xtheta}).
The explicit expressions of the first four $\widetilde F_k$'s 
in terms of the partition functions $\widetilde Z_k$ are
\begin{equation}
 \begin{aligned}
  \widetilde F_1 &=  \lim_{\cE\to 0}\, \cE \,\widetilde Z_1~,\\
\widetilde F_2 &=  \lim_{\cE\to 0}\, \cE\big(\widetilde Z_2-
\frac{1}{2}\,\widetilde Z_1^2\big)~,\\
\widetilde F_3 &=  \lim_{\cE\to 0}\, \cE\big(\widetilde Z_3-
\widetilde Z_2\widetilde Z_1
+\frac{1}{3}\,\widetilde Z_1^3\big)~,\\
\widetilde F_4&=  \lim_{\cE\to 0}\, \cE\big(\widetilde Z_4-
\widetilde Z_3\widetilde Z_1
-\frac{1}{2}\widetilde Z_2^2+\widetilde Z_2\widetilde Z_1^2
-\frac{1}{4}\,\widetilde Z_1^4\big)~.
 \end{aligned}
\label{tildeFks}
\end{equation}
Given the prepotential (\ref{tildeFk}) we obtain the corresponding non-perturbative source current for the axio-dilaton from \eq{stquantum}, namely
\begin{equation}
\label{Jnp}
J_{\mathrm{n.p.}} = -\frac{(2\pi\alpha')^4}{2\pi}\,\bar\delta \widetilde F_{\mathrm{n.p.}}
= -\frac{(2\pi\alpha')^4}{2\pi} \sum_{k=1}^\infty q^k\, \bar\delta \widetilde F_k~,
\end{equation}
where $\bar\delta$ is the variation operator defined in \eq{bardelta}. 
Exploiting the Leibniz rule and \eq{zk00}, we easily obtain
\begin{equation}
 \begin{aligned}
 \bar\delta \widetilde F_1 &=  \lim_{\cE\to 0}\, \cE \,\bar\delta \widetilde Z_1~,\\
\bar\delta \widetilde F_2 &=  \lim_{\cE\to 0} \,\cE\big(\bar\delta \widetilde Z_2-
Z_1\,\bar\delta\widetilde Z_1\big)~,\\
\bar\delta \widetilde F_3 &=  \lim_{\cE\to 0} \,\cE\big(\bar\delta\widetilde Z_3-
Z_1\,\bar\delta\widetilde Z_2- Z_2\,\bar\delta\widetilde Z_1
+Z_1^2\,\bar\delta\widetilde Z_1\big)~,\\
\bar\delta \widetilde F_4&=  \lim_{\cE\to 0}\, \cE\big(\bar\delta\widetilde Z_4-
Z_1\,\bar\delta\widetilde Z_3-Z_3\,\bar\delta\widetilde Z_1
-Z_2\,\bar\delta\widetilde Z_2+Z_1^2\,\bar\delta\widetilde Z_2
+2 Z_2 Z_1\,\bar\delta\widetilde Z_1
-Z_1^3\,\bar\delta\widetilde Z_1\big)~.
 \end{aligned}
\label{bardeltatildeFk}
\end{equation}
Using the explicit form of the instanton action $\widetilde S_{\mathrm{inst}}$ 
given in \eq{sinst1}, it is rather straightforward to show that
\begin{equation}
 \bar\delta\widetilde Z_k = 4\pi\ii \sum_{\ell=0}^\infty 
(2\pi\alpha')^{2\ell}\,\bar p^{2\ell+4}\,Z_k^{(2\ell)}
\label{bardeltaZ}
\end{equation}
where we have defined
\begin{equation}
 Z_k^{(2\ell)} = \frac{1}{(2\ell)!}\,\int d\cM_{(k)} \,\tr(\chi^{2\ell})\,
\ee^{-\widetilde S_{\mathrm{inst}}}\,\Big|_{T=\tau_0\,,\,M=m_{\mathrm{cl}}}~.
\label{Zkell}
\end{equation}
Notice that
\begin{equation}
\label{zk0}
 Z_k^{(0)}= k\,Z_k~.
\end{equation}
Furthermore, since for $k=1$ there are no $\chi$ moduli, we simply have
\begin{equation}
 Z_1^{(2\ell)} = 0~~~\forall \,\ell\not=0~,
\label{Z1}
\end{equation}
so that the following relation holds: $\bar\delta\widetilde Z_1 = 4\pi\ii \,\bar p^{4}\,Z_1$.

Equipped with this information, we can rewrite \eq{bardeltatildeFk} as follows
\begin{equation}
 \begin{aligned}
 \bar\delta \widetilde F_1 &=  4\pi\ii\,\bar p^4\Big\{\lim_{\cE\to 0}\, \cE Z_1\Big\}~,\\
\bar\delta \widetilde F_2 &=  4\pi\ii \sum_{\ell=0}^\infty 
(2\pi\alpha')^{2\ell}\,\bar p^{2\ell+4}\Big\{\lim_{\cE\to 0}\,\cE\big(Z_2^{(2\ell)}-
Z_1Z_1^{(2\ell)}\big)\Big\}~,\\
\bar\delta \widetilde F_3 &=  4\pi\ii \sum_{\ell=0}^\infty 
(2\pi\alpha')^{2\ell}\,\bar p^{2\ell+4}\Big\{\lim_{\cE\to 0} \,\cE\big(Z_3^{(2\ell)}-
Z_1Z_2^{(2\ell)}- Z_2Z_1^{(2\ell)}
+Z_1^2Z_1^{(2\ell)}\big)\Big\}~,\\
\bar\delta \widetilde F_4&= 4\pi\ii \sum_{\ell=0}^\infty 
(2\pi\alpha')^{2\ell}\,\bar p^{2\ell+4}\Big\{ \lim_{\cE\to 0}\, \cE\big( Z_4^{(2\ell)}-
Z_1Z_3^{(2\ell)}-Z_3Z_1^{(2\ell)}
-Z_2Z_2^{(2\ell)}\\
&~~~~~~~~~~~~~~~~~~~~~~~~~~~~~~~~~~~~~~~~~~~~+Z_1^2Z_2^{(2\ell)}+2 Z_2 Z_1Z_1^{(2\ell)}
-Z_1^3Z_1^{(2\ell)}\big)\Big\}~.
 \end{aligned}
\label{bardeltatildeFk1}
\end{equation}
We now show that the expressions in braces are directly related to the elements of the chiral ring
of the SO(8) gauge theory living on the D7 brane world-volume.

\subsection{The SO(8) chiral ring}
\label{subsec:ringSO8}
The elements of the chiral ring of the SO(8) gauge theory are the vacuum expectation values
of the traces of powers of the adjoint scalar field from the D7/D7 strings, 
{\it i.e.} $\langle\tr m^J\rangle$.
The non-perturbative contributions to these correlators can be computed via localization methods
by inserting suitable operators in the instanton partition function as shown in 
Ref.~\cite{Fucito:2009rs}. Specifically we have
\begin{equation}
\begin{aligned}
\langle\tr m^J\rangle_{\mathrm{n.p.}} &=\sum_{k=1}^\infty q^k\langle\tr m^J\rangle_k\\
&=
\lim_{\cE\to 0}\Big\{\frac{1}{\widetilde{\mathcal{Z}}}\,
\sum_{k=1}^\infty q^k\!\int \!d\cM_{(k)}\,\ee^{-\widetilde S_{\mathrm{inst}}}\,\cO_{(k,J)}\Big\}
\Bigg|_{T=\tau_0\,,\,M=m_{\mathrm{cl}}}
\end{aligned}
\label{mJ}
\end{equation}
where
\begin{equation}
\begin{aligned}
\cO_{(k,J)} = \sum_{I=1}^k\Big[&\chi_I^J - \sum_{i}(\chi_I+\varepsilon_i)^J 
+\sum_{i<j}(\chi_I+\varepsilon_i+\varepsilon_j)^J\\
&-\sum_{i<j<\ell}(\chi_I+\varepsilon_i+\varepsilon_j+\varepsilon_\ell)^J 
+(\chi_I+\varepsilon_1+\varepsilon_2+\varepsilon_3+\varepsilon_4)^J \Big]~.
\end{aligned}
\label{OkJ}
\end{equation}
Here the $\chi_I$'s are the elements of the $\chi$ matrix when it is written in a skew-diagonalized
form, {\it i.e.} in the Cartan basis of SO($k$). More explicitly we have
\begin{itemize}
 \item[$\bullet$] for $k=2n$ \begin{equation}  
   \big\{\chi_I\big\} = \big\{\chi_1,\cdots,\chi_n,-\chi_1,\cdots,-\chi_n\big\}~,
                             \end{equation}
\item[$\bullet$] for $k=2n+1$ \begin{equation}  
   \big\{\chi_I\big\} = \big\{\chi_1,\cdots,\chi_n,-\chi_1,\cdots,-\chi_n,\,0\big\}~,
                             \end{equation}
\end{itemize}
and
\begin{equation}
\label{trchiell}
 \tr(\chi^{2\ell}) = \sum_{I=1}^k (-1)^\ell\,\chi_I^{2\ell}~.
\end{equation}
Due to the symmetry of the theory, all odd elements of the chiral ring vanish, {\it i.e.}
\begin{equation}
 \langle \tr m^J \rangle_{\mathrm{n.p.}}  = 0 ~~~\mbox{for}~J~\mbox{odd}~.
\label{phiodd}
\end{equation}
Furthermore, it is immediate to check that $\cO_{(k,2)}=0$ for all $k$,
so that
\begin{equation}
 \langle \tr m^2 \rangle_{\mathrm{n.p.}}  = 0 ~.
\label{m2np}
\end{equation}
Therefore, the first non-trivial element of the SO(8) chiral ring is 
$\langle \tr m^4 \rangle_{\mathrm{n.p.}}$, which now we are going to analyze in some detail.

Setting $J=4$ in \eq{OkJ}, we easily find
\begin{equation}
 \label{cO4}
\cO_{(k,4)}= 24\,\cE\,k~;
\end{equation}
from this it follows that
\begin{equation}
\begin{aligned}
 \langle\tr m^4\rangle_{\mathrm{n.p.}} &= 24\,
\lim_{\cE\to 0}\Big\{\frac{\cE}{\widetilde{\mathcal{Z}}}\,
\sum_{k=1}^\infty k\,q^k \widetilde{\mathcal{Z}}_k\Big\}
\Bigg|_{T=\tau_0\,,\,M=m_{\mathrm{cl}}}
\\ &
= 24\,q\,\frac{\partial}{\partial q}\,\lim_{\cE\to 0}
\Big\{\cE \log \widetilde{\mathcal{Z}}\Big\}
\Bigg|_{T=\tau_0\,,\,M=m_{\mathrm{cl}}} \\
&= 24\,q\,\frac{\partial}{\partial q}\,\widetilde F_{\mathrm{n.p.}}\Big|_{T=\tau_0\,,\,M=m_{\mathrm{cl}}} 
\end{aligned}
 \label{m4}
\end{equation}
where in the last step we introduced the prepotential according to \eq{tildeF}. 
Expanding in powers of $q$ and using \eq{tildeFks}, at the first few instanton numbers we find
\begin{equation}
 \begin{aligned}
\langle\tr m^4\rangle_1 &=  24\widetilde F_1\Big|_{T=\tau_0\,,\,M=m_{\mathrm{cl}}}
=  24\,\lim_{\cE\to 0}\, \cE Z_1~,\\
\langle\tr m^4\rangle_2 &=  48\widetilde F_2\Big|_{T=\tau_0\,,\,M=m_{\mathrm{cl}}} 
= 24\,\lim_{\cE\to 0}\,\cE\big(2Z_2-Z_1^2\big)~,\\
\langle\tr m^4\rangle_3 &=  72\widetilde F_3\Big|_{T=\tau_0\,,\,M=m_{\mathrm{cl}}} 
= 24\,\lim_{\cE\to 0}\,\cE \big(3Z_3-3Z_2Z_1
+Z_1^3\big)~,\\
\langle\tr m^4\rangle_4 &=  96\widetilde F_4\Big|_{T=\tau_0\,,\,M=m_{\mathrm{cl}}} 
= 24\,\lim_{\cE\to 0}\,\cE \big(4Z_4-4Z_3Z_1-2Z_2^2+4Z_2Z_1^2
-Z_1^4\big)~.
 \end{aligned}
\label{m4k}
\end{equation}
Recalling that $Z_k^{(0)}=k\,Z_k$, in the right hand sides above we recognize 
the same expressions of the $\ell=0$ terms of the quantities appearing inside braces in \eq{bardeltatildeFk1}.

These results can be generalized order by order in the instanton expansion to the higher elements of the SO(8) chiral ring.
We refer to Appendix \ref{app:chiral} for the details while here we simply quote the results 
for the first few instanton numbers. For any integer $\ell>0$ we find
\begin{equation}
 \begin{aligned}
\langle\tr m^{2\ell+4}\rangle_1 &=  0~,\\
\langle\tr m^{2\ell+4}\rangle_2 &=  (-1)^{\ell}\,(2\ell+4)!\,
\lim_{\cE\to 0}\,\cE Z_2^{(2\ell)}~,\\
\langle\tr m^{2\ell+4}\rangle_3 &=  (-1)^{\ell}\,(2\ell+4)!\,\lim_{\cE\to 0}\,\cE 
\big(Z_3^{(2\ell)}-Z_1Z_2^{(2\ell)}\big)~,\\
\langle\tr m^{2\ell+4}\rangle_4 &=  (-1)^{\ell}\,(2\ell+4)!\,\lim_{\cE\to 0}\,\cE 
\big( Z_4^{(2\ell)}-Z_1Z_3^{(2\ell)}-Z_2Z_2^{(2\ell)}
+Z_1^2Z_2^{(2\ell)}\big)~.
 \end{aligned}
\label{mlk}
\end{equation}
Since $Z_1^{(2\ell)}=0$ for $\ell>0$, the right hand sides have precisely the same structures
appearing in \eq{bardeltatildeFk1}. 

The non-perturbative correlators $\langle\tr m^{J}\rangle_k$ have been computed for
the first few values of $J$ and $k$ in Ref.~\cite{Fucito:2009rs}; here, for completeness 
but also as an illustration of how explicit these computations are,
we report their results in our current conventions%
\footnote{Notice that the instanton expansion parameter $q$ of Ref.~\cite{Fucito:2009rs} is mapped 
to our $-q$. 
Notice also that our current mass eigenvalues $m_i$ are rescaled with a factor of $1/\sqrt{2}$ with respect to those used in Ref.~\cite{Billo:2010mg}, {\it i.e.} $m_i^{\mathrm{here}}
=m_i^{\mathrm{there}}/\sqrt{2}$.}, namely
\begin{equation}
\begin{aligned}
 \label{chirexp}
 \vev{\tr m^2}_{\mathrm{n.p.}} & = 0\phantom{\Big(}~,\\
 \vev{\tr m^4}_{\mathrm{n.p.}} & = - 192\, 
\Pf m ~q -96\sum_{i<j} m_i^2 m_j^2 ~ q^2 -768\, \Pf m~ q^3 + \ldots~,\\
 \vev{\tr m^6}_{\mathrm{n.p.}} & =  1440\sum_{i<j<k} m_i^2 m_j^2
 m_k^2 ~q^2 + 7680\, \Pf m\, \sum_i m_i^2~ q^3 + \ldots ~,\\
 \vev{\tr m^8}_{\mathrm{n.p.}} & = - 6720\, (\Pf m)^2 ~q^2 - 35840\,\Pf m\,
 \sum_{i<j}m_i^2 m_j^2~ q^3 + \ldots ~.
\end{aligned}
\end{equation}
where $\Pf m=m_1m_2m_3m_4$.

\subsection{The $\tau$ profile}
\label{subsec:taunp}

Using the results (\ref{m4k}) and (\ref{mlk}), the coefficients 
$\bar\delta\widetilde F_k$ of prepotential can be
rewritten in a compact way as follows
\begin{equation}
 \bar\delta\widetilde F_k = 4\pi\ii \sum_{\ell=0}^\infty 
(-1)^\ell\,\frac{(2\pi\alpha')^{2\ell}}{(2\ell+4)!}\,\bar p^{2\ell+4}
\,\vev{\tr m^{2\ell+4}}_k~.
\end{equation}
Thus, after summing over all instanton numbers $k$ and suitably relabeling the index $\ell$, the non-perturbative source current (\ref{Jnp}) becomes
\begin{equation}
 \label{Jnpfin}
J_{\mathrm{n.p.}} = -2\ii \sum_{\ell=1}^\infty (-1)^\ell\,
\frac{(2\pi\alpha')^{2\ell}\,\vev{\tr m^{2\ell}}_{\mathrm{n.p.}}}{(2\ell)!} \,{\bar p}^{2\ell}~.
\end{equation}
Here we have used the fact that $\vev{\tr m^2}_{\mathrm{n.p.}} = 0$ to set the starting value
of the summation index at $\ell=1$. This is exactly of the form (\ref{Jquantumft}) 
expected from the general analysis of Section \ref{sec:dcd7}. {From} the
current (\ref{Jnpfin}) we deduce the following non-perturbative axio-dilaton profile
\begin{equation}
\label{taunp}
\widetilde\tau_{\mathrm{n.p.}}(z) = - \frac{1}{2\pi\ii}\sum_{\ell=1}^\infty
\frac{(2\pi\alpha')^{2\ell}}{2\ell}\,\frac{\vev{\tr m^{2\ell}}_{\mathrm{n.p.}}}{z^{2\ell}}~.
\end{equation}
This analysis shows the existence of a direct connection between the elements of the
SO(8) chiral ring and the non-perturbative source terms for the axion-dilaton field
that was already observed in Ref.~\cite{Billo:2010mg}.

Introducing the quantity
\begin{equation}
 a= \frac{z}{2\pi\alpha'}
\label{az}
\end{equation}
which in the holographic correspondence parametrizes the Coulomb branch of the dual
$\cN=2$ SU(2)$\,\sim\,$Sp(1) gauge theory, and adding to $\widetilde\tau_{\mathrm{n.p.}}$ 
the classical contribution (\ref{massexpm}), we can exhibit the resulting axio-dilaton $\tau$
as an expansion in inverse powers of $a$, namely
\begin{eqnarray}
  2\pi\ii\tau & =& 2\pi\ii\tau_0 - \frac{\sum_i m_i^2}{a^2} + 
  \frac{1}{a^4} \!\left[\!-\frac 12 \sum_i m_i^4 + 48\, \Pf m \,q 
  + 24 \!\sum_{i<j} m_i^2 m_j^2 ~q^2 + 192 \,\Pf m\,q^3 + \!\ldots\right] \nonumber\\
  && + \frac{1}{a^6} \left[\!- \frac{1}{3} \sum_i m_i^6 
  - 240 \sum_{i<j<k}m_i^2 m_j^2 m_k^2 ~ q^2 - 1280\, \Pf m\, \sum_i m_i^2~ q^3 +
  \ldots \right]\label{taua}\\
  &&+ \frac{1}{a^8} \!\left[\!-\frac{1}{4} \sum_i m_i^8 + 
  840\, (\Pf m)^2~ q^2
  + 4480\, \Pf m\, \sum_{i<j} m_i^2 m_j^2
  ~ q^3 + \ldots\right]+~\ldots~.
\nonumber
\end{eqnarray}
As shown in Ref.~\cite{Billo:2010mg}, this is in perfect agreement with the 
effective coupling derived from the SW curve for the SU(2) 
$N_f=4$ theory \cite{Seiberg:1994aj}.

\section{An orbifold example}
\label{sec:orbifold}
The techniques described in the previous sections are very flexible and
can be applied in different contexts; in particular they do not require an eight-dimensional set-up. 
To show this we now briefly analyze an orbifold model that yields results directly
in four-dimensions.
\subsection{Moduli spectrum and effective action}
\label{subsec:orbifold}
We consider the type I$^\prime$ theory described in Section \ref{sec:dcd7} and mod it out
with a $\mathbb Z_2$ projection generated by
\begin{equation}
\label{orbact}
 g~:~x^m~\rightarrow~-x^m ~~~~\mbox{for}~~m=4,\ldots,7~.
\end{equation}
The ``Lorentz'' group $\mathrm{SO}(8)$ considered so far is then broken to 
$\mathrm{SO}(4)\times \mathrm{SO}(4)$, and all representations 
split correspondingly. For example, the eight-dimensional vector index decomposes into two 
four-dimensional indices, which we denote respectively by $\mu=0,\ldots,3$ and $m=4,\ldots,7$ (as already indicated in \eq{orbact}), one for each SO(4) factor, and so on. In the spinor representation
$g$ can be identified by the chirality operator of the second SO(4), 
so that objects carrying a chiral (or anti-chiral) spinor index $a$ (or $\dot a$) of this SO(4) are 
even (or odd) under the orbifold. 

In this model tadpole cancellation requires the presence of two fractional D7 branes 
extending both along the four space-time directions $x^\mu$ and along the orbifolded ones $x^m$. 
If the Chan-Paton factors of the D7 branes do not transform under $g$ (that is,
if the D7 branes are in the trivial $\mathbb Z_2$ representation), the low-energy open string
excitations on the D7 branes give rise to a theory with gauge group SO(4) as shown in Ref.~\cite{Fucito:2009rs}.

The non-perturbative sectors of this theory are obtained by adding fractional D(--1) branes.
The corresponding instanton moduli are easily obtained by projecting the spectrum described in
Section \ref{sec:dcmoduli} onto the sub-space of moduli that are even under the $\mathbb Z_2$ action. 
In the neutral Neveu-Schwarz sector we find the vector $a^\mu$ and the 
two scalars $\chi$ and $\bar{\chi}$, 
since the $a^m$'s are projected out. In the neutral Ramond sector the surviving moduli are
$\eta_{\alpha a}$ and $\lambda_{\dot{\alpha}a}$, where the index $\alpha$ ($\dot\alpha$) labels
the chiral (anti-chiral) spinor representation of the first SO(4), while $\eta_{\dot \alpha \dot a}$
and $\lambda_{\alpha\dot a}$ are projected out.
Also in this orbifold model, the consistency with the orientifold projection requires that $a^\mu$ and $\eta_{\alpha a}$ transform in the symmetric representation of the instanton symmetry group 
SO$(k)$, while $\chi,\bar{\chi}$ and $\lambda_{\dot{\alpha}a}$ must transform in the antisymmetric 
representation. Again the abelian part of $a^\mu$ and $\eta_{\alpha a}$ can be identified with
the supercoordinates of a chiral superspace, this time in four dimensions, namely
\begin{equation}
\label{xthetaorb}
 x_\mu \sim \ell_s^2 \,\tr(a_\mu)\quad,\quad\theta_{\alpha a}\sim \ell_s^2 \,\tr (\eta_{\alpha a})
\end{equation}
in complete analogy with \eq{xtheta}.
In the charged sector, the fermionic moduli $\mu$, transforming in the bi-fundamental representation of $\mathrm{SO}(k)\times \mathrm{SO}(4)$, are unaffected by the orbifold projection 
and thus remain in the spectrum.

The instanton action in this orbifold model can be simply obtained by projecting the one described
in Section \ref{sec:dcmoduli} on the even moduli. Since this projection does not change the
formal structure of the action, we do not write it again. However it is worth mentioning
that now there is a non-vanishing coupling between the axion $\widetilde C_0$ and four 
$\theta$'s instead of eight, as it is explicitly shown in Appendix \ref{subapp:C0theta}
where the disk amplitude among the axion vertex and four $\theta$ vertices is computed.
This result gives a direct evidence that the axio-dilaton field $\tau$ 
can be promoted to a full $\cN=2$ chiral superfield $T$ as follows
\begin{equation}
\label{T44}
T = \tau_0 + \widetilde T = \tau_0 +\widetilde\tau + \sqrt{2}\theta \widetilde\lambda + ... + 
\theta^4 \frac{\partial^2~}{\partial z^2}\widetilde{\overline\tau}~. 
\end{equation} 
Therefore, we can conclude that the complete moduli action 
has the same form given in Eq.s~(\ref{sinst}) and 
(\ref{sinst1}) with $\widetilde T$ given by \eq{T44} instead of \eq{Tsuper}.

\subsection{The non-perturbative axio-dilaton source}
In this orbifold model the D-instantons induce a non-perturbative four-dimensional
effective action given by
\begin{equation}
 S_{\mathrm{n.p.}} = \frac{1}{(2\pi)^2}\int\! d^4x \,d^4\theta\,\widetilde F_{\mathrm{n.p.}}, 
\label{effact5}
\end{equation}
where the prepotential $\widetilde F_{\mathrm{n.p.}}$ is defined 
as in \eq{tildeFk} but with the integrals performed over the centered moduli described 
in the previous subsection. Correspondingly, 
the non-perturbative source term for the axio-dilaton is  
\begin{equation}
 J_{\mathrm{n.p.}}=-\frac{(2\pi\alpha')^2}{2\pi}\,\bar\delta \widetilde F_{\mathrm{n.p.}}
=-\frac{(2\pi\alpha')^2}{2\pi}\, \frac{\delta\widetilde 
F_{\mathrm{n.p.}}}{\delta\big(\theta^4\,\widetilde{\overline\tau}\big)}
\Bigg|_{T=\tau_0\,,\,M=m_{\mathrm{cl}}}
\label{jnporb}
\end{equation}
where now $m_{\mathrm{cl}}=\mathrm{diag}\{m_1,m_2,-m_1,-m_2\}$. Expanding in powers of $q$, 
one finds that that the coefficients $\bar{\delta}\widetilde{F}_k$ are formally equal 
to the those written in \eq{bardeltatildeFk}, but with $\bar{\delta}\widetilde{Z}_k$
given by
\begin{equation}
\label{Z4}
\bar{\delta}\widetilde{Z}_k=-2\pi\ii \sum_{\ell=0}^\infty (2\pi\alpha')^{2\ell}\bar{p}^{2\ell+2}Z_k^{(2\ell)}.
\end{equation}
In particular the terms up to $k=4$ are
\begin{equation}
\begin{aligned}
\bar\delta \widetilde F_1 &=  -2\pi\ii\,\bar p^2\Big\{\lim_{\cE\to 0}\, \cE Z_1\Big\}~,
\\
\bar\delta \widetilde F_2 &=  -2\pi\ii 
\sum_{\ell=0}^\infty (2\pi\alpha')^{2\ell}\,\bar p^{2\ell+2}
\Big\{\lim_{\cE\to 0}\,\cE\big(Z_2^{(2\ell)}-Z_1Z_1^{(2\ell)}\big)\Big\}~,
\\
\bar\delta \widetilde F_3 &=  -2\pi\ii 
\sum_{\ell=0}^\infty (2\pi\alpha')^{2\ell}\,\bar p^{2\ell+2}
\Big\{\lim_{\cE\to 0} \,\cE\big(Z_3^{(2\ell)}-Z_1Z_2^{(2\ell)}- Z_2Z_1^{(2\ell)}
+Z_1^2Z_1^{(2\ell)}\big)\Big\}~,
\\
\bar\delta \widetilde F_4&= -2\pi\ii 
\sum_{\ell=0}^\infty (2\pi\alpha')^{2\ell}\,\bar p^{2\ell+2}
\Big\{ \lim_{\cE\to 0}\, \cE\big( Z_4^{(2\ell)}-Z_1Z_3^{(2\ell)}-Z_3Z_1^{(2\ell)}
-Z_2Z_2^{(2\ell)}\\
&~~~~~~~~~~~~~~~~~~~~~~~~~~~~~~~~~~~~~~~~~~~~+Z_1^2Z_2^{(2\ell)}+2 Z_2 Z_1Z_1^{(2\ell)}
-Z_1^3Z_1^{(2\ell)}\big)\Big\}~.
 \end{aligned}
\label{bdtFk14}
\end{equation}

The right hand sides can be rewritten in terms of the chiral ring 
of the SO(4) theory that has been computed using Nekrasov's prescription 
and localization techniques in Ref.~\cite{Fucito:2009rs}. 
The non-perturbative SO(4) correlators $\langle\tr m^J\rangle_{\mathrm{n.p.}}$
are as in \eq{mJ} but with $\mathcal{O}_{(k,J)}$ given by
\begin{equation}
\label{OkJorb}
\cO_{(k,J)}= \sum_{I=1}^k\Big[\chi_I^J - (\chi_I+\varepsilon_1)^J -
(\chi_I+\varepsilon_2)^J 
+(\chi_I+\varepsilon_1+\varepsilon_2)^J\Big]~.
\end{equation}
Explicit computations show that non-trivial contributions arise only for $J$ 
even due to the symmetry of the theory. The first non-trivial element, corresponding to $J=2$, 
can be expressed in a closed form for any instanton number $k$ since $\cO_{(k,2)}= 2\,\cE\,k$, where 
$\cE=\varepsilon_1\varepsilon_2$. Therefore we have
\begin{equation}
\begin{aligned}
 \langle\tr m^2\rangle_{\mathrm{n.p.}} &= \sum_{k=1}^\infty q^k\langle\tr m^2\rangle_k= 2\,
\lim_{\cE\to 0}\Big\{\frac{\cE}{\widetilde{\mathcal{Z}}}\,
\sum_{k=1}^\infty k\,q^k \widetilde{\mathcal{Z}}_k\Big\}
\Bigg|_{T=\tau_0\,,\,M=m_{\mathrm{cl}}}
\\ &
= 2\,q\,\frac{\partial}{\partial q}\,\lim_{\cE\to 0}
\Big\{\cE \log {\mathcal{Z}}\Big\} =
2\,q\,\frac{\partial}{\partial q}\,F_{\mathrm{n.p.}}~.
\end{aligned}
 \label{m2orb}
\end{equation}
The subsequent elements of the chiral ring for $J>2$ can be explicitly computed order by order
in the instanton expansion with exactly the same methods described in Section \ref{sec:profile}
and Appendix \ref{app:chiral}. The results for the first few instanton numbers and $\ell>0$
are
\begin{equation}
 \begin{aligned}
\langle\tr m^{2\ell+2}\rangle_1 &=  0~,\\
\langle\tr m^{2\ell+2}\rangle_2 &=  (-1)^{\ell}\,(2\ell+2)!\,
\lim_{\cE\to 0}\,\cE Z_2^{(2\ell)}~,\\
\langle\tr m^{2\ell+2}\rangle_3 &=  (-1)^{\ell}\,(2\ell+2)!\,\lim_{\cE\to 0}\,\cE 
\big(Z_3^{(2\ell)}-Z_1Z_2^{(2\ell)}\big)~,\\
\langle\tr m^{2\ell+2}\rangle_4 &=  (-1)^{\ell}\,(2\ell+2)!\,\lim_{\cE\to 0}\,\cE 
\big( Z_4^{(2\ell)}-Z_1Z_3^{(2\ell)}-Z_2Z_2^{(2\ell)}
+Z_1^2Z_2^{(2\ell)}\big)~.
 \end{aligned}
\label{m44k}
\end{equation}
Since $Z_1^{(2\ell)}=0$ for $\ell>0$, the right hand sides have precisely the same structures
appearing in \eq{bdtFk14}.
We thus can write
\begin{equation}
 \bar\delta\widetilde F_k = -2\pi\ii \sum_{\ell=0}^\infty 
(-1)^\ell\,\frac{(2\pi\alpha')^{2\ell}}{(2\ell+2)!}\,\bar p^{2\ell+2}
\,\vev{\tr m^{2\ell+2}}_k~,
\end{equation}
so that \eq{jnporb} becomes
\begin{equation}
J_{\mathrm{n.p.}}=
\ii \sum_{\ell=0}^\infty 
(-1)^\ell\,\frac{(2\pi\alpha')^{2\ell+2}}{(2\ell+2)!}\,\bar p^{2\ell+2}
\,\vev{\tr m^{2\ell+2}}_{\mathrm{n.p.}}~.
\end{equation}
For completeness we report, in our conventions, 
the explicit expressions of $\langle\tr m^J\rangle_{\mathrm{n.p.}}$
for the first few values of $J$ and the first few instanton numbers, that have been computed in Ref.~\cite{Fucito:2009rs}, namely
\begin{equation}
\begin{aligned}
\vev{\tr m^2}_{\mathrm{n.p.}} & = 
4\,\Pf m \,q-2\sum_i m_i^2 \,q^2+4\,\Pf m \,q^3-2\sum_i m_i^2\, q^4+\ldots,\phantom{\Big(}~,\\
 \vev{\tr m^4}_{\mathrm{n.p.}} & = - 12\, 
(\Pf m)^2 \,q^2 +16\,\Pf m\,\sum_{i} m_i^2 \,q^3-6\Big( \!\big(\sum_i m_i^2\big)^2
+6\,(\Pf m)^2\Big)q^4
+ \ldots~,\\
\vev{\tr m^6}_{\mathrm{n.p.}} & =40\,(\Pf m)^3\,q^3-90\,(\Pf m)^2 \,\sum_i m_i^2 \,q^4+\ldots,\\
\langle\tr m^8\rangle_{\text{n.p.}}& =-140\,(\Pf m)^4 \,q^4+\ldots
\end{aligned}
\end{equation}
where $\Pf m=m_1m_2$.
 
\section{Conclusions}
\label{sec:conclusions}
The conclusion we can draw is that the non-perturbative corrections to the profile of closed string fields representing the gravitational dual of gauge couplings can be microscopically derived by including the modification of the source terms in their equations of motion induced by instantonic branes. We have explicitly shown this non-perturbative holographic correspondence in the 
particular case of the axio-dilaton in Sen's local limit of the type I$^\prime$ 
theory recovering the complete F-theory background, and also in an orbifolded version of it.
Remarkably, the non-perturbative corrections to the gauge couplings 
can be entirely reconstructed from the knowledge of the chiral ring of the flavor degrees of freedom.

It would be extremely interesting, and we think it should be possible, to generalize this derivation 
to other cases and to other gravitational fields.
\vfill \eject

\noindent {\large {\bf Acknowledgments}}
\vskip 0.2cm
We thank  F. Fucito, L. Gallot, J.F. Morales and I. Pesando
for several very useful discussions. 
 
\vskip 1cm
\appendix

\section{Notations and conventions}
\label{app:notconv}
We denote the $10d$ space coordinates by $x^M$, $M=0,\ldots 9$, those along
the world-volume of the D7 branes by $x^\mu$, while for the last two directions
we introduce the complex coordinates 
\begin{equation}
\label{defz}
z=x^8 - \ii \,x^9~,~~~
\bar z=x^8 + \ii \,x^9~.
\end{equation}
Thus, our brane set-up breaks the $10d$ ``Lorentz'' group 
$\mathrm{SO}(10)$ into $\mathrm{SO}(8)\times \mathrm{SO}(2)$. 
We will indicate the bosonic string fields corresponding to the coordinates (\ref{defz})
as $Z$ and $\overline Z$.
In the complex basis, the non-vanishing metric components read $g_{z\bar z} =
g_{\bar z z} = 1/2$; of course then $g^{z\bar z} = g^{\bar z z} = 2$. We
indicate by 
\begin{equation} 
\label{defholder} \partial = \frac 12 \big(\partial_8 +
\ii\, \partial_9\big)~,~~~ \bar\partial = \frac 12 \big(\partial_8 - \ii \,\partial_9\big)
\end{equation} 
the holomorphic, respectively anti-holomorphic, derivatives. We also introduce 
the complex combination of momenta
\begin{equation}
\label{defcmom}
p = \frac 12 \big(p_8 - \ii p_9\big)~,~~~
\bar p = \frac 12 \big(p_8 + \ii p_9\big)~.
\end{equation}
The scalar product $\vec p\cdot \vec x$ in the last two directions can then be expressed as $\bar pz + p\bar z$.

The Laplace operator in these directions can be written as $\square =
4\bar\partial\partial$. We define the two-dimensional $\delta$-function
$\delta^2(x^8,x^9)$, which we sometimes loosely denote as $\delta^2(z)$, with
respect to the integration measure $dx^8\, dx^9$, so that the logarithm
satisfies the Laplace equation with the following normalization
\begin{equation} 
\label{lapeq} \square \log \big(|z|^2/|z_0|^2\big) = 4\pi\, \delta^2(z)~.
\end{equation}

At the level of the Clifford algebra, if we denote by $\widehat\Gamma^M$ the
32-dimensional $\gamma$-matrices for SO$(10)$, we have the decomposition
\begin{equation}
 \label{decogamma}
\widehat\Gamma^\mu = \Gamma^\mu \otimes \one~,~~~
\widehat\Gamma^8 = \Gamma \otimes \sigma^1~,~~~
\widehat\Gamma^9 = \Gamma \otimes \sigma^2~,
\end{equation}
where $\Gamma^\mu$ are the 16-dimensional $\gamma$-matrices for SO(8), and
$\Gamma$ the corresponding chirality matrix. They are tensored with matrices
acting on the 2-dimensional spinor space for SO(2).
In the complex basis for the last two directions we have
in particular
\begin{equation}
\label{decozzb}
\widehat\Gamma^z = \widehat\Gamma^8 - \ii \,\widehat\Gamma^9 = 2 \Gamma \otimes 
\begin{pmatrix}
\,0\, & \,0\, \\
\,1\, & \,0\,
\end{pmatrix}~,~~~
\widehat\Gamma^{\bar z} = \widehat\Gamma^8 + \ii \,\widehat\Gamma^9 = 2 \Gamma \otimes
\begin{pmatrix}
 \,0\, & \,1\, \\
\,0\, & \,0\,
\end{pmatrix}~;
\end{equation}
using the metric in the complex basis to lower the indices, we also have
$\hat\Gamma_z = \frac 12 \hat\Gamma^{\bar z}$ and $\hat\Gamma_{\bar z} = \frac
12 \hat\Gamma^{z}$.

The $10d$ chirality matrix $\hat\Gamma$ decomposes as
\begin{equation}
 \label{Gammadec}
\widehat\Gamma = \Gamma \otimes \sigma^3~.
\end{equation}
If we denote the chiral (or antichiral) spinor indices of $\mathrm{SO}(10)$ by
$\cA$ (or $\dot\cA$), those of $\mathrm{SO}(8)$ by $\alpha$ (or $\dot\alpha$)
and those of $\mathrm{SO}(2)$ by $+$ (or $-$), this decomposition corresponds to
splitting these indices as follows
\begin{equation}
\label{splitspi}
\cA = (\alpha,+) \cup (\dot\alpha,-)~,~~~
\dot\cA = (\dot\alpha,+) \cup (\alpha,-)~.
\end{equation}
Thus, if $\Theta^{\dot\cA}$ is a $10d$ anti-chiral Majorana-Weyl spinor, it
decomposes as
\begin{equation}
\label{decoTheta}
\Theta^{\dot\cA} \to (\bar\theta^{\dot\alpha +},\theta^{\alpha -})~;
\end{equation}
if the $10d$ anti-chiral nature of these spinor is known, the $\pm$ indices become
redundant, and we often do not write them.

The $10d$ charge conjugation matrix $\widehat{\mathcal{C}}$, such that 
$\widehat{\cC}\,\widehat\Gamma^M\,{\widehat{\cC}}^{-1} = -\,{}^t\widehat\Gamma^M$, becomes
\begin{equation}
\label{decocharge}
\widehat{\cC} = -\ii\, \cC\otimes \sigma^2~,
\end{equation}
where $\cC$ is the charge conjugation matrix in $8d$, which satisfies 
$\cC\,\Gamma^M\,\cC^{-1} = -\,{}^t\Gamma^M$, while the matrix $\sigma^2$ 
represents charge conjugation in the last two dimensions.

For the SO(8) Clifford algebra it is possible to choose a chiral basis
in which
\begin{equation}
\label{chir8}
\Gamma^\mu = \begin{pmatrix}
0 & \gamma^\mu \\ \bar\gamma^\mu & 0
\end{pmatrix}
~,~~~
\Gamma = \begin{pmatrix}
\one_8 & 0 \\ 0 & -\one_8
\end{pmatrix}
~,~~~\cC = \begin{pmatrix}
C & 0\\
0 & C
\end{pmatrix}~,
\end{equation}
with $C = \sigma^2\otimes \sigma^3\otimes\sigma^2$
when written as a tensor product of two-dimensional factors.
In this basis the spinor indices $\alpha$ and $\dot\alpha$ enumerate a specific ordering of the spinor weights. This is particularly well suited to deal with world-sheet 
computations involving spin fields, whose charges in the bosonized description are exactly the components of the spinor weights.

The massless vertex in the Ramond-Ramond sector of the closed superstring,
in the $(-\frac 12,-\frac 12)$ picture, reads
\begin{equation}
\label{RRver}
\pi g_s \ell_s\,F_{\dot\cA\dot\cB}\, S^{\dot\cA}(w)\,
\ee^{\ii\,\ell_s p_L\cdot X(w)} \,\ee^{-\frac{1}{2}\varphi(w)}~
{\widetilde S}^{\dot\cB}(\bar w)\,
\ee^{\ii\,\ell_s p_R\cdot\widetilde X(\bar w)}\,\ee^{-\frac{1}{2}\widetilde\varphi(\bar w)}~.
\end{equation}
Here $S^{\dot\cA}$ and ${\widetilde S}^{\dot\cB}$ are $10d$ left and right spin-fields
while $\ee^{-\frac{1}{2}\varphi}$ and $\ee^{-\frac{1}{2}\widetilde\varphi}$ are the superghost factors; our conventions are that the type IIB GSO projection singles out antichiral spinors in both left- and 
right-moving sectors. The bispinor polarization $F_{\dot\cA\dot\cB}$, when
expanded in $p$-forms, reads
\begin{equation}
\label{expFgamma}
F_{\dot\cA\dot\cB} = \frac 18 \sum_{p~\mathrm{odd}} F_{M_1\ldots M_p}
\,\big(\cC\widehat\Gamma^{M_1\ldots M_p}\big)_{\dot\cA\dot\cB}~,
\end{equation}
where $F_{M_1\ldots M_p}$ is the field-strength of the Ramond-Ramond $(p-1)$-form potential.

Consider the Ramond-Ramond scalar $\widetilde C_0$, depending only on the last two directions;
its field strength is a 1-form has the non-vanishing component $F_z = \partial_z \widetilde C_0$ 
({\it i.e.} $\ii\,\bar p\,\widetilde C_0$ in momentum space).
It follows from Eq.s (\ref{RRver}) and (\ref{expFgamma}) that the corresponding
vertex contains the matrix
\begin{equation}
\label{cgz}
\widehat{\cC}\,\widehat\Gamma^z = -2\left(\begin{matrix}
\cC\Gamma & 0 \\
0 & 0
\end{matrix}\right)
\end{equation}
which, using \eq{chir8}, has the only non-vanishing components
\begin{equation}
\label{cgsp}
(\widehat{\cC}\widehat\Gamma^z)_{\alpha +\,,\,\beta +} =
-2 C_{\alpha\beta}~,~~~
(\widehat{\cC}\widehat\Gamma^z)_{\dot\alpha +\,,\,\dot\beta +} =
2 C_{\dot\alpha\dot\beta}~.
\end{equation}
The second set of these components is antichiral in $10d$ and appears thus in axion
vertex (\ref{rr}).

The notations for vector and spinor indices in the orbifold case considered in Section \ref{sec:orbifold} are given in the main text below \eq{orbact}. Let us just remark that in this case we replace the spinor index $\alpha$ of $\mathrm{SO}(8)$ with a couple $(\alpha,a)$ or $(\dot\alpha,\dot a)$, where $\alpha$ and $a$ are chiral spinor indices for the first and the second factors in the decomposition $\mathrm{SO}(8)\to \mathrm{SO}(4)\times \mathrm{SO}(4)$;
similarly for the antichiral case.

\section{Relevant string diagrams}
\label{app:sd}
This appendix contains some details about the calculation of disk diagrams that
produce the couplings of the axio-dilaton to the D-instantons moduli considered in the
main text.

\subsection{Interaction among $\widetilde C_0$ and the $\theta$ moduli}
\label{subapp:C0theta}
Here we describe the explicit computation of the coupling among the Ramond-Ramond scalar 
$\widetilde C_0$ and the $\theta$-moduli, corresponding to the higher term 
in the closed superfield $T$.

\subsubsection*{Orbifold case} 
Let us first consider the set-up described in Section \ref{sec:orbifold}, where
this coupling is technically easier to derive, with respect to the type I'
set-up, since it contains only four $\theta$ insertions.

We consider thus a disk diagram having D$(-1)$ boundary conditions and 
insert an axion vertex in the interior and four vertices for the fermionic moduli 
$\eta$ on the boundary
 \begin{equation}
 \label{ampli0}
 A = \lvev V_\eta V_\eta V_\eta V_\eta V_{\widetilde C_0}\rvev~. 
 \end{equation}
Actually, we are interested in the components of the fermionic moduli
along the identity (see \eq{xtheta}), namely
 \begin{equation}
 \label{etatothetan}
  \eta_{\alpha a} = \cN\, \ell_s^{-2}\, \theta_{\alpha a}\, \one_k~,
\end{equation}
where the notation for spinor indices is the one introduced in Section \ref{subsec:orbifold}, and
$\cN$ is a purely numerical factor that we will fix soon.
We take the axion vertex in the $(-1/2, -1/2)$ picture, so it has the form of \eq{rr}.
Since on the disk the total (right + left) superghost charge must equal $-2$, three open 
string vertices, say those with polarizations $\eta_{\alpha_i a_i}$ located at positions $x_i$ with $i=1,2,3$, can be taken in the $(- 1/2)$ picture, as in Table \ref{tab:mod}, while the fourth 
one has to be taken in the $(1/2)$ picture; the latter has the form
\begin{equation}
\label{vertpum}
\ell_s^{3/2} \,\eta_{\alpha_4 a_4}\, \partial \bar Z(x_4)\, 
S^{\alpha_4 a_4 +}(x_4)\,\ee^{+\frac{1}{2}\varphi(x_4)}~.
\end{equation}

Since we are inserting the maximum number of $\theta$'s, the spinor indices of the
four fermionic vertices have to be all different. We can thus choose a fixed value for 
them, and then sum over all cyclically inequivalent orderings. For instance,
we can make the choice indicated below:
\begin{equation}
\label{tabspf}
\begin{tabular}{|c|c|c|c|}
\hline
&$\phi_1\ldots\phi_4$&$\phi_5$&$\varphi$\\
\hline
$\alpha_1a_1-$&$++++$&$-$&$-$\\
$\alpha_2a_2-$&$----$&$-$&$-$\\
$\alpha_3a_3-$&$++--$&$-$&$-$\\
$\alpha_4a_4+$&$--++$&$+$&$+$\\
$\dot\cA$&$+---$&$+$&$-$\\
$\dot\cB$&$-+++$&$+$&$-$\\
\hline
\end{tabular}
\end{equation}
The last column reports (twice) the superghost charge of the various vertices.
Moreover, $\phi_1,\ldots \phi_5$ are the world-sheet scalars that bosonize the spin
fields in the five couples of directions. The spinor indices are associated to
(twice) their charges with respect to these fields, {\it i.e.} to twice the 
corresponding weight vectors. 
We have also taken into account the fact that the spinor indices
$\dot\cA$ and $\dot\cB$ of the spin fields of the axion vertex must be such 
as to saturate all charges;
their last component must thus be of type $+$, while their first four components
must take values opposite to each other. Different choices of these values (we have altogether 8 possibilities) lead
to contributions that would arise from diagrams with different orderings of the
$\theta$'s, so we can fix them as in the table
above and multiply the result by 8.

Taking into account the above considerations, the disk amplitude (\ref{ampli0})
becomes
\begin{equation}
\label{ampli1}
A=\frac{2\pi}{g_s} ~2\pi g_s \ell_s~ \ii\, \bar p\, \widetilde C_0~ \cN^4 \,\ell_s^{-2}
\, k \,
\theta^4\, 
\int\frac{\prod_{i=1}^4dx_i\,dw\,d\bar{w}}{dV_{\mathrm{CKG}}}
C_{\mathrm{tot}}~.
\end{equation}
Here the factor $\frac{2\pi}{g_s}$ is the topological normalization of the disk,
the remaining overall constants arise from the normalizations and polarizations
of the vertices, while $dV_{\mathrm{CKG}}$ stands for the volume of the conformal Killing group.
With $C_{\mathrm{tot}}$ we have denoted the sum over the inequivalent ordering
of the four open string vertices, that is 
\begin{equation}
\label{suminord}
C_{\mathrm{tot}} = C_{1234} + C_{2134} + C_{3214} + C_{1324} + C_{2314} + C_{3124}
\end{equation}
where
\begin{equation}
\label{ampli2}
C_{1234} = S_{1234}(x_i,w,\bar{w}) X(x_4,w,\bar{w}).
\end{equation}
Here $S,X$ are the relevant correlators of spin fields and bosonic world-sheet fields,
computed with the choice of indices in \eq{tabspf}. After identifying the left and right
components of the closed string vertex ($\widetilde S^{\dot{\cB}}=S^{\dot{\cB}}$
and $\widetilde Z=-Z$), one finds
\begin{eqnarray}
S_{1234}(x_i,w,\bar{w}) & =&
\big\langle S^{\alpha_1a_1-}(x_1)\,S^{\alpha_2a_2-}(x_2)\,S^{\alpha_3a_3-}(x_3)\,
S^{\alpha_4a_4+}(x_4)\,S^{\dot{\cA}}(w)\,S^{\dot{\cB}}(\bar{w})\big\rangle\nonumber\\
&&~~~~\phantom{\Big|}\big\langle\ee^{-\frac{1}{2}\varphi(x_1)}\,\ee^{-\frac{1}{2}\varphi(x_2)}
\,\ee^{-\frac{1}{2}\varphi(x_3)}\,\ee^{+\frac{1}{2}\varphi(x_4)}\,\ee^{-\frac{1}{2}\varphi(w)}
\ee^{-\frac{1}{2}\varphi(\bar w)}\big\rangle
\label{s1234}\\
& =& \big[(x_1-x_2)(x_1-w)(x_2-\bar{w})(x_3-x_4)(x_3-\bar{w})(w-\bar{w})\big]^{-1}(x_4-\bar{w})
\nonumber
\end{eqnarray}
and
\begin{equation} 
\label{Xcorr}
X(x_4,w,\bar{w}) =\big\langle\partial\overline{Z}(x_4)\,
\ee^{\ii\,\ell_s\bar p  Z(w)}\,\ee^{-\ii\,\ell_s\bar p Z(\bar w)}\big\rangle
 = \ii\ell_s\,\bar{p}\,\frac{(w-\bar{w})}{(x_4-w)(x_4-\bar{w})}~,
\end{equation}
so that altogether
\begin{equation}
\label{c1234is}
C_{1234}=\ii\ell_s\bar{p}\,\big[(x_1-x_2)(x_1-w)(x_2-\bar{w})
(x_3-x_4)(x_3-\bar{w})(x_4-w)\big]^{-1}.
\end{equation}

The other inequivalent correlators contributing to $C_{\mathrm{tot}}$ can be
computed by inserting the vertices with indices $\alpha_i a_i$ at permuted locations.
Each of them can be equivalently represented in four different ways obtained by cyclically permuting the four open string vertices, which is guaranteed to lead to the same result after integration. It turns out to be convenient, rather than using \eq{suminord}, to rewrite 
$C_{\mathrm{tot}}$ as one-half the sum of twelve terms, two in each equivalence class, chosen so that  some partial sums algebraically simplify. For instance, one has
\begin{equation}
 \begin{aligned}
\label{4cex}
 C_{1234}&=(\ii\ell_s\bar{p})\,\big[(x_1-x_2)(x_1-w)(x_2-\bar{w})(x_3-x_4)(x_3-\bar{w})
(x_4-w)\big]^{-1}~,\\
 C_{1243}&=(-\ii\ell_s\bar{p})\,\big[(x_1-x_2)(x_1-w)(x_2-\bar{w})(x_3-x_4)(x_4-\bar{w})
(x_3-w)\big]^{-1}~,\\
 C_{2143}&=(\ii\ell_s\bar{p})\,\big[(x_1-x_2)(x_2-w)(x_1-\bar{w})(x_3-x_4)(x_3-\bar{w})
(x_4-w)\big]^{-1}~,\\
 C_{2134}&=(-\ii\ell_s\bar{p})\,\big[(x_1-x_2)(x_2-w)(x_1-\bar{w})(x_3-x_4)(x_4-\bar{w})
(x_3-w)\big]^{-1}~,
\end{aligned}
\end{equation}
so that
\begin{equation}
\label{partC1}
C_{(1)} = C_{1234}+C_{1243}+C_{2143}+C_{2134}=(-\ii\ell_s\bar{p})\frac{(w-\bar{w})^2}{\prod_{i=1}^4\lvert x_i-w\rvert^2}~.
\end{equation}
Similarly, one finds
\begin{equation}
\label{partC2}
C_{(2)} = 
C_{1324}+C_{1423}+C_{2314}+C_{2413}= C_{(1)}
\end{equation}
and
\begin{equation}
\label{partC3}
C_{(3)} = C_{1342}+C_{1432}+C_{2341}+C_{2431}= C_{(1)}~.
\end{equation}
The total correlator  arising from the sum over inequivalent orderings is thus
\begin{equation}
C_{\text{tot}}=\frac{1}{2}\left(C_{(1)} + C_{(2)} + C_{(3)}\right) = 
\frac 32
(-\ii\ell_s\bar{p})\frac{(w-\bar{w})^2}{\prod_{i=1}^4\lvert x_i-w\rvert^2}~.
\end{equation}
We can now insert this expression into \eq{ampli1} and proceed to evaluate the integral.

Fixing $x_1\rightarrow\infty,w=\ii,\bar{w}=-\ii$, the integrand gets multiplied by $\lvert x_1-w\rvert^2(w-\bar{w})$ and the integral becomes
\begin{multline}
\label{risint}
-\frac{3\ii\ell_s\bar{p}}{2}(2\ii)^3\int_{-\infty}^{+\infty}dx_2\int_{-\infty}^{x_2}dx_3\int_{-\infty}^{x_3}dx_4\prod_{i=2}^4(x_i^2+1)^{-1}\\
=-\frac{3\ii\ell_s\bar{p}}{2}(2\ii)^3\frac{1}{3!}\int_{-\infty}^{+\infty}\prod_{i=2}^4dx_i\prod_{i=2}^4(x_i^2+1)^{-1}=-\frac{(2\pi)^3}{4}\ell_s\bar{p}~.
\end{multline}
Substituting this result into \eq{ampli1} we get finally the amplitude 
\begin{equation}
\label{finampl}
A = -2\pi \ii\, k\, \frac{(2\pi)^4\, \cN^4}{4} \bar p^2 \widetilde C_0\, \theta^4~.
\end{equation}
If we choose
\begin{equation}
\label{Nchoice}
\cN = \frac{\sqrt{2}}{2\pi}~,
\end{equation}
this amplitude corresponds to the following term in the instanton moduli action:
\begin{equation}
\label{efampl}
-2\pi \ii \,k\, \theta^4\,\partial^2\widetilde C_0~.
\end{equation}
This coupling promotes the Ramond-Ramond part of $-2\pi\ii k\widetilde\tau$ appearing 
in the moduli action (see \eq{S01a}) to the highest Ramond-Ramond part of 
$-2\pi\ii k\widetilde T$ where
\begin{equation} 
\label{supT4}
\widetilde T = \widetilde \tau + \sqrt{2} \theta \widetilde\lambda + 
\ldots + \theta^4 \frac{\partial^2}{\partial z^2}\widetilde{\bar\tau}~,
\end{equation}
represents the analogous of the type I$^\prime$ closed string superfield of \eq{Tsuper}
in the orbifold case.

\subsection*{The flat case}
The previous calculation easily generalizes to the local limit of type I$^\prime$ theory 
considered in Sections \ref{sec:dcd7}-\ref{sec:profile}, where the disk diagram describing the interaction of the axion $\widetilde C_0$ with the maximum number of $\theta$s, that in this case is eight, can again be computed by exactly the same kind of techniques, namely by
fixing explicit values for the spinor indices for one ordering and summing over non-equivalent re-orderings.
In this case, five of the $\theta$ vertices are in the $(-1/2)$ picture and three in the 
$(+1/2)$ picture, so that from the bosonic correlator we get two extra powers of $\bar p$ with respect to the previous computation.
All in all, taking into account the different combinatorics, the result is that
\begin{equation}
\label{ampli8theta}
A  = \lvev V_\eta V_\eta V_\eta V_\eta V_\eta V_\eta V_\eta V_\eta V_{\widetilde C_0}\rvev
= 2\pi \ii\, k\, \frac{(2\pi)^8\, \cN^8}{8} \bar p^4 \widetilde C_0\, \theta^8~.
\end{equation}
Using \eq{Nchoice}, this corresponds to a term in the moduli action of the form
\begin{equation}
\label{efamplf}
-2\pi \ii \,k\, 2\theta^8\,\partial^4\widetilde C_0~,
\end{equation}
which is indeed part of the interaction term $-2\pi\ii k\widetilde T$ 
involving the closed string superfield of \eq{Tsuper} (see Eq.s (\ref{S01b}) and
(\ref{theta8bartau}).

\subsection{Interaction of $\widetilde C_0$ with many $\chi$-moduli}
\label{subapp:C0chi}
Here we consider the interaction among the axion $\widetilde C_0$ and any 
(even) number $n$ of $\chi$ moduli, namely
\begin{equation}
\label{chiC00}
A=\lvev \underbrace{V_{\chi}\cdots V_\chi}_{n=2\ell}\, V_{\widetilde C_0} \rvev~,
\end{equation}
and sketch the derivation of its expression, reported in \eq{chiC0},
which was crucial in our discussion. For this computation no substantial difference arises in considering the flat or the orbifolded case, so we proceed directly in the former case.

Again the axion vertex is taken in the picture $(-\frac{1}{2},-\frac{1}{2})$, and is given in \eq{rr}. One of the $\chi$ vertices, which we place in position $x_1$, is in the superghost picture $(-1)$ 
and is thus given in Table \ref{tab:mod}, while all the others are in the picture $(0)$, namely 
\begin{equation}
\label{vchi0}
V_\chi^{(0)}(x_i) = \ell_s \,\chi\, \partial\overline Z(x_i)~.
\end{equation}
Then the amplitude (\ref{chiC00}) takes the form 
\begin{equation}
\label{AchiC0}
\begin{aligned}
A =&\frac{2\pi}{g_s} \,\frac{2\pi g_s\ell_s}{8} \,(\ell_s)^{n}\, 
\tr(\chi^n)\, \ii\,\bar p\, \widetilde C_0\, \frac{(n-1)!}{n!} \\
& \times 
\int\frac{\prod_{i=1}^ndx_i\,dw\,d\bar{w}}{dV_{\mathrm{CKG}}}
A_1(x_1,w,\bar{w})\,A_2(x_1,w,\bar{w})\,A_3(x_i,w,\bar{w})~.
\end{aligned}
\end{equation}
This expression includes the disk normalization factor $2\pi/g_s$ and
polarizations and normalizations of the vertices; there is an $1/n!$ symmetry
factor but a $(n-1)!$ arises from the sum over cyclically inequivalent orderings
of the vertices, since it easy to see that all of them lead to the same answer.
Finally, $A_{1,2,3}$ are respectively the superghost, spin fields and bosonic correlators:
\begin{align}
A_1(x_1,w,\bar{w})& =
\big\langle \ee^{-\varphi(x_1)}\,\ee^{-\frac{1}{2}\varphi(w)}\,
\ee^{-\frac{1}{2}\varphi(\bar{w})}
\big\rangle~,\\
A_2(x_1,w,\bar{w})& = \cC_{\dot\alpha\dot\beta}
\,\big\langle\overline{\Psi}(x_1)\,S^{\dot\alpha+}(w)\,S^{\dot\alpha+}(\bar{w})\big\rangle~,\\
A_3(x_i,w,\bar{w})& =
\big\langle\partial\bar{Z}(x_2)\ldots\partial\bar{Z}(x_n)\ee^{\ii\,\ell_s\bar p  Z(w)}\,\ee^{-\ii\,\ell_s\bar p Z(\bar w)}\big\rangle~.
\end{align}
Using standard methods we obtain 
\begin{align}
A_1(x_1,w,\bar{w})A_2(x_1,w,\bar{w})& = 8\,\big[(x_1-w)(x_1-\bar{w})(w-\bar{w})\big]^{-1},\\
A_3(x_i,w,\bar{w})& =(-\ii\ell_s\bar{p})^{n-1}\frac{(w-\bar{w})^{n-1}}{\prod_{i=2}^n\lvert x_i-w\rvert^2}.
\end{align}
Fixing $x_1\rightarrow\infty,w=\ii,\bar{w}=-\ii$, the integrand gets multiplied by 
$\lvert x_1-w\rvert^2(w-\bar{w})$ and the integral becomes
\begin{equation}
\begin{split}
& (-\ii\ell_s\bar{p})^{n-1}(2\ii)^{n-1}\int_{-\infty}^{+\infty}dx_2\int_{-\infty}^{x_2}dx_3\ldots\int_{-\infty}^{x_{n-1}}dx_n\prod_{i=2}^n(x_i^2+1)^{-1}\\
& =2^{n-1}\ell_s^{n-1}\bar{p}^{n-1}\frac{1}{(n-1)!}\int_{-\infty}^{+\infty}\prod_{i=2}^ndx_i\prod_{i=2}^n(x_i^2+1)^{-1}\\
& =\frac{(2 \pi\ell_s)^{n-1}}{(n-1)!}\bar{p}^{n-1}.
\end{split}
\end{equation}
The amplitude reads therefore
\begin{equation}
\label{finresA}
A = 2\pi\ii\, (2\pi\alpha')^n \,\frac{\tr(\chi^{n})}{n!}\, \bar p^n\, \widetilde C_0~,
\end{equation}
where we have taken into account that $\ell_s = \sqrt{\alpha'}$. This is just \eq{chiC0} 
used in the main text.

\section{D7 brane chiral ring and non-perturbative axio-dilaton sources}
\label{app:chiral}
In this Appendix we give some further details on the D-instanton induced source
terms for the axio-dilaton and their relation with the elements of the chiral
ring on the D7 branes.
The non-perturbative current $J_{\mathrm{n.p.}}$ is given in \eq{Jnp} in terms
of the quantities $\widetilde \delta F_k$, which in turn are expressed in
\eq{bardeltatildeFk1} in terms of the instanton partition functions $Z_k$ and of
the quantities $Z_k^{(2\ell)}$ defined in \eq{Zkell}. 

It is possible to show that the expressions appearing in \eq{bardeltatildeFk1}
are directly related to the non-perturbative contributions to the elements
$\vev{\tr m^J}_{\mathrm{n.p.}}$ of the D7 brane chiral ring. This relation,
which represents one of the main points of Section \ref{sec:profile}, 
has been explicitly shown there in the case of $\vev{\tr m^4}_{\mathrm{n.p.}}$.
Here we give further details on how this relation arises.

As shown in Ref.~\cite{Fucito:2009rs}, $\vev{\tr m^J}_{\mathrm{n.p.}}$ can be
computed via localization techniques through the formula given in \eq{mJ},
which we rewrite here follows:
\begin{equation}
\langle\tr m^J\rangle_{\mathrm{n.p.}} =\sum_{k=1}^\infty q^k\langle\tr
m^J\rangle_k = \lim_{\cE\to 0} \Big\{\frac{1}{\cZ}\,
\sum_{k=1}^\infty q^k Z_{(k,J)}\Big\}~,
\label{mJapp}
\end{equation}
where
\begin{equation}
 \label{defZjk}
Z_{(k,J)} = \int \!d\cM_{(k)}\,\ee^{- S_{\mathrm{inst}}}\,\cO_{(k,J)}~,
\end{equation}
with $\cO_{(k,J)}$ given in \eq{OkJ}. Notice that the right hand side contains the moduli
action evaluated at $T=\tau_0\,,\, M=m_{\mathrm{cl}}$, which we distinguish by
stripping off the tilde sign. Analogously, $\cZ$ is the instanton partition
function of \eq{tildeZ} evaluated at $T=\tau_0\,,\, M=m_{\mathrm{cl}}$.

Comparing the coefficient of $q^k$ in \eq{mJapp} we have, for the first few
instanton levels, 
\begin{equation}
 \begin{aligned}
 \langle \tr m^J \rangle_{1} &= \lim_{\cE\to 0} Z_{(1,J)}~,\\
\langle \tr m^J \rangle_{2} &=\lim_{\cE\to 0}
\big(Z_{(2,J)}-Z_1\,Z_{(1,J)}\big)~,\\
\langle \tr m^J \rangle_{3} &=\lim_{\cE\to 0}
\big(Z_{(3,J)}-Z_1\,Z_{(2,J)}
-Z_2\,Z_{(1,J)}+Z_1^2\,Z_{(1,J)}\big)~,\\
\langle \tr m^J \rangle_{4} &=\lim_{\cE\to 0} \big(
Z_{(4,J)}-Z_1\,Z_{(3,J)}-Z_2\,Z_{(2,J)}+Z_1^2\,Z_{(2,J)}\\
&~~~~~~~~~~~~+2Z_2\,\,Z_1\,Z_{(1,J)}-Z_3\,Z_{(1,J)}-Z_1^3\,Z_{(1,J)}\big)~,
 \end{aligned}
\label{phiJk}
\end{equation}
and so on. In Section \ref{sec:profile} we argued that $\langle\tr
m^J\rangle_{\mathrm{n.p.}}=0$ for $J$ odd, so the next non-trivial case to be considered 
after the cases $J=2,4$ given in the main text, is $J=6$. 
{From} \eq{OkJ} it is possible to show that
\begin{equation}
 \label{Ok6}
\cO_{(k,6)} =
60\,k\,\cE\,h_2(\varepsilon_i)-360\,\cE\,{\mathrm{tr}}(\chi^2)~,
\end{equation}
where we have defined
\begin{equation}
h_2(\varepsilon_i)= 2\sum_{i}\varepsilon_i^2 + 3\sum_{i<j} \varepsilon_i\varepsilon_j~.
\label{h2}
\end{equation}
Therefore, it follows that the integrals $Z_{(k,J)}$ are related to the
partition functions $Z_k$ and to the integrals with $\chi$ insertions
$Z_k^{(2\ell)}$ defined in \eq{Zkell}, according to
\begin{equation}
 Z_{(k,6)} = 60\,k\,\cE\,h_2(\varepsilon_i)\,Z_k - 720\,\cE\, Z_k^{(2)}~.
\label{Zk6}
\end{equation}
Substituting this result into \eq{phiJk}, we obtain (taking into account that
$Z_1^{(2\ell)}=0$ for all $\ell$) 
\begin{align}
& \langle \tr m^6 \rangle_{1} = \lim_{\cE\to 0}
60\,h_2(\varepsilon_i)\,\cE\,Z_1 = 
\lim_{\cE\to 0} 60\,h_2(\varepsilon_i)\,F_1~,
\label{Z16}\\
&\langle \tr m^6 \rangle_{2}  = \lim_{\cE\to 0} \Big\{
 -720\,\cE\,Z_2^{(2)} 
+120\,h_2(\varepsilon_i)\,\cE\big(Z_2-\frac{1}{2}\,Z_1^2\big)\Big\} \nonumber\\
&~~~~~~~~~~\,=\lim_{\cE\to 0} \Big\{ -720\,\cE\,Z_2^{(2)}
+120\,h_2(\varepsilon_i)\,\cE\big(F_2 + \cO(\varepsilon_i)\big)\Big\}~,
\label{z26}\\
&\langle \tr m^6 \rangle_{3} = \lim_{\cE\to 0} \Big\{
 -720\,\cE\,\big(Z_3^{(2)}- Z_1\,Z_2^{(2)}\big)
+180\,h_2(\varepsilon_i)\,\cE\big(Z_3-Z_2 Z_1+\frac{1}{3}\,Z_1^3\big)\Big\}\nonumber\\
 &~~~~~~~~~~\,= \lim_{\cE\to 0} \Big\{-720\,\cE\,\big(Z_3^{(2)}-
Z_1\,Z_2^{(2)}\big)+180\,h_2(\varepsilon_i)\,\cE\big(F_3 +
\cO(\varepsilon_i)\big)\Big\}~,
\end{align}
\begin{align}
&\langle \tr m^6 \rangle_{4}  = \lim_{\cE\to 0} \Big\{
-720\,\cE\,\big(Z_4^{(2)}-Z_1\,Z_3^{(2)}
-Z_2\,Z_2^{(2)}+Z_1^2\,Z_2^{(2)}\big) \nonumber\\
&\quad\quad\quad~\quad\quad\quad\quad+240\,h_2(\varepsilon_i)\,\cE\big(Z_4-Z_3\,Z_1-\frac{1}{2}
\,Z_2^2 +Z_2\,Z_1^2-\frac{1}{4}\,Z_1^4\big)\Big\}\nonumber\\
 &~~~~~~~~~~\,= \lim_{\cE\to 0} \Big\{ 
-720\,\cE\,\big(Z_4^{(2)}-Z_1\,Z_3^{(2)}
-Z_2\,Z_2^{(2)}+Z_1^2\,Z_2^{(2)}\big)\nonumber\\
&\quad\quad\quad~\quad\quad\quad\quad+ 240\,h_2(\varepsilon_i)\,\cE\,\big(F_4 +
\cO(\varepsilon_i)\big)\Big\}~.
\end{align}
We see that all terms in these expressions which contain only the ordinary instanton partition functions $Z_k$ arrange themselves in the combinations $F_k$. These are the coefficients in the instanton expansion of the non-perturbative prepotential introduced in Eq.s (\ref{tildeF})-(\ref{tildeFks}), evaluated at $T=\tau_0\,,\, M=m_{\mathrm{cl}}$.
The prepotential is finite in the $\cE\to 0$ limit, so that the above equations reduce to
\begin{equation}
 \begin{aligned}
\langle \tr m^6 \rangle_{2} &=-6!\,\lim_{\cE\to 0} \cE\,Z_2^{(2)}~,\\
\langle \tr m^6 \rangle_{3} &=-6!\,\lim_{\cE\to 0} \cE\,\big(Z_3^{(2)}-
Z_1\,Z_2^{(2)}\big)~,\\
\langle \tr m^6 \rangle_{4} &=-6!\,\lim_{\cE\to 0} \cE\,\big(Z_4^{(2)}-Z_1\,Z_3^{(2)} -Z_2\,Z_2^{(2)}+Z_1^2\,Z_2^{(2)}\big)~,
 \end{aligned}
\label{phi6234}
\end{equation}
and so on. We recognize in the right hand sides above exactly the same expressions appearing at the order $\ell=3$ in the variations $\delta{\widetilde F_k}$ as given in \eq{bardeltatildeFk1}. 

Let us now consider the third non-trivial element of the chiral ring, namely $\langle \tr m^8 \rangle_{\mathrm{n.p.}}$. {From} \eq{OkJ}, one can show with a bit of work that
\begin{equation}
\label{o8j}
\cO_{(k,8)} = 56\,k\,\cE\,h_4(\varepsilon_i)-3360\,\cE\,h_2(\epsilon_i)\,\tr(\chi^2)
+8!\,\cE\,\tr(\chi^4)~,
\end{equation}
where we have defined
\begin{equation}
h_4(\varepsilon_i)= 6\sum_{i}\varepsilon_i^2 + 3\sum_{i\not =j} \varepsilon^3_i\varepsilon_j
+20\sum_{i<j}\varepsilon_i^2\varepsilon_j^2 +30 \sum_{i<j}\sum_{l\not =i,j}\varepsilon_i\varepsilon_j\varepsilon_l^2
+45\,\cE~.
\label{h4}
\end{equation}
It follows that 
\begin{equation}
 Z_{(k,8)} =56\,k\,\cE\,h_4(\varepsilon_i)\,Z_k-3360\,\cE\,h_2(\varepsilon_i)\,Z_k^{(2)}
+8!\,\cE\,Z_k^{(4)}~.
\label{Zk8}
\end{equation}
Inserting this expression into \eq{phiJk} we again find that all terms containing 
only $Z_k$'s arrange themselves in the prepotential coefficients $F_k$ and disappear in the limit $\cE\to 0$; moreover also the terms containing $Z_k^{(2)}$ vanish in this limit so that 
we remain only with 
\begin{equation}
 \begin{aligned}
\langle \tr m^8 \rangle_{2} &=8!\,\lim_{\cE\to 0} \cE\,Z_2^{(4)}~,\\
\langle \tr m^8 \rangle_{3} &=8!\,\lim_{\cE\to 0} \cE\,\big(Z_3^{(4)}-
Z_1\,Z_2^{(4)}\big)~,\\
\langle \tr m^8 \rangle_{4} &=8!\,\lim_{\cE\to 0} \cE\,\big(Z_4^{(4)}-Z_1\,Z_3^{(4)} -Z_2\,Z_2^{(4)}+Z_1^2\,Z_2^{(4)}\big)~,
 \end{aligned}
\label{phi8234}
\end{equation}
and so on. This corresponds to the $\ell=4$ terms in the right hand side of \eq{bardeltatildeFk1}.

We infer that this correspondence holds for every value of $\ell$, as written in \eq{mlk} in the main text, and links the non-perturbative source terms for the axio-dilaton, hence its profile, to the elements of the chiral ring on the D7 branes.

\providecommand{\href}[2]{#2}\begingroup\raggedright\endgroup


\begin{thebibliography}{10}

\bibitem{Maldacena:1997re}
J.~M. Maldacena, \emph{{The large N limit of superconformal field theories and
  supergravity}}, \href{http://dx.doi.org/10.1023/A:1026654312961,
  10.1023/A:1026654312961}{Adv.Theor.Math.Phys. {\bf 2} (1998)  231--252},
  \href{http://arxiv.org/abs/hep-th/9711200}{{\tt arXiv:hep-th/9711200
  [hep-th]}}.

\bibitem{Gubser:1998bc}
S.~S. Gubser, I.~R. Klebanov, and A.~M. Polyakov, \emph{{Gauge theory
  correlators from non-critical string theory}},
  \href{http://dx.doi.org/10.1016/S0370-2693(98)00377-3}{Phys. Lett. {\bf B428}
  (1998)  105--114},
\href{http://arxiv.org/abs/hep-th/9802109}{{\tt arXiv:hep-th/9802109}}.

\bibitem{Witten:1998qj}
E.~Witten, \emph{{Anti-de Sitter space and holography}}, Adv. Theor. Math.
  Phys. {\bf 2} (1998)  253--291,
\href{http://arxiv.org/abs/hep-th/9802150}{{\tt arXiv:hep-th/9802150}}.

\bibitem{Polyakov:1998ju}
A.~M. Polyakov, \emph{{The wall of the cave}},
  \href{http://dx.doi.org/10.1142/S0217751X99000324}{Int.J.Mod.Phys. {\bf A14}
  (1999)  645--658}, \href{http://arxiv.org/abs/hep-th/9809057}{{\tt
  arXiv:hep-th/9809057 [hep-th]}}.

\bibitem{Klebanov:2000hb}
I.~R. Klebanov and M.~J. Strassler, \emph{{Supergravity and a confining gauge
  theory: Duality cascades and chi SB resolution of naked singularities}}, JHEP
  {\bf 0008} (2000)  052, \href{http://arxiv.org/abs/hep-th/0007191}{{\tt
  arXiv:hep-th/0007191 [hep-th]}}.

\bibitem{Maldacena:2000yy}
J.~M. Maldacena and C.~Nunez, \emph{{Towards the large N limit of pure N=1
  superYang-Mills}},
  \href{http://dx.doi.org/10.1103/PhysRevLett.86.588}{Phys.Rev.Lett. {\bf 86}
  (2001)  588--591}, \href{http://arxiv.org/abs/hep-th/0008001}{{\tt
  arXiv:hep-th/0008001 [hep-th]}}.

\bibitem{Seiberg:1994rs}
N.~Seiberg and E.~Witten, \emph{{Monopole condensation and confinement in N=2
  supersymmetric Yang-Mills Theory}},
  \href{http://dx.doi.org/10.1016/0550-3213(94)90124-4}{Nucl. Phys. {\bf B426}
  (1994)  19--52},
\href{http://arxiv.org/abs/hep-th/9407087}{{\tt arXiv:hep-th/9407087}}.

\bibitem{Seiberg:1994aj}
N.~Seiberg and E.~Witten, \emph{{Monopoles, duality and chiral symmetry
  breaking in N=2 supersymmetric QCD}},
  \href{http://dx.doi.org/10.1016/0550-3213(94)90214-3}{Nucl. Phys. {\bf B431}
  (1994)  484--550},
\href{http://arxiv.org/abs/hep-th/9408099}{{\tt arXiv:hep-th/9408099}}.

\bibitem{Klebanov:1999rd}
I.~R. Klebanov and N.~A. Nekrasov, \emph{{Gravity duals of fractional branes
  and logarithmic RG flow}},
  \href{http://dx.doi.org/10.1016/S0550-3213(00)00016-X}{Nucl.Phys. {\bf B574}
  (2000)  263--274}, \href{http://arxiv.org/abs/hep-th/9911096}{{\tt
  arXiv:hep-th/9911096 [hep-th]}}.

\bibitem{Bertolini:2000dk}
M.~Bertolini, P.~Di~Vecchia, M.~Frau, A.~Lerda, I.~Pesando, and R.~Marotta,
  \emph{{Fractional D-branes and their gauge duals}}, JHEP {\bf 02} (2001)
  014,
\href{http://arxiv.org/abs/hep-th/0011077}{{\tt arXiv:hep-th/0011077}}.

\bibitem{Polchinski:2000mx}
J.~Polchinski, \emph{{N=2 Gauge / gravity duals}},
  \href{http://dx.doi.org/10.1142/S0217751X01003834}{Int.J.Mod.Phys. {\bf A16}
  (2001)  707--718}, \href{http://arxiv.org/abs/hep-th/0011193}{{\tt
  arXiv:hep-th/0011193 [hep-th]}}.

\bibitem{Bertolini:2001qa}
M.~Bertolini, P.~Di~Vecchia, M.~Frau, A.~Lerda, and R.~Marotta, \emph{{N = 2
  gauge theories on systems of fractional D3/D7 branes}},
  \href{http://dx.doi.org/10.1016/S0550-3213(01)00568-5}{Nucl. Phys. {\bf B621}
  (2002)  157--178},
\href{http://arxiv.org/abs/hep-th/0107057}{{\tt arXiv:hep-th/0107057}}.

\bibitem{Billo:2001vg}
M.~Billo, L.~Gallot, and A.~Liccardo, \emph{{Classical geometry and gauge duals
  for fractional branes on ALE orbifolds}},
  \href{http://dx.doi.org/10.1016/S0550-3213(01)00399-6}{Nucl.Phys. {\bf B614}
  (2001)  254--278}, \href{http://arxiv.org/abs/hep-th/0105258}{{\tt
  arXiv:hep-th/0105258 [hep-th]}}.

\bibitem{DiVecchia:2005vm}
P.~Di~Vecchia, A.~Liccardo, R.~Marotta, and F.~Pezzella, \emph{{On the gauge /
  gravity correspondence and the open/closed string duality}},
  \href{http://dx.doi.org/10.1142/S0217751X05024900}{Int. J. Mod. Phys. {\bf
  A20} (2005)  4699--4796},
\href{http://arxiv.org/abs/hep-th/0503156}{{\tt arXiv:hep-th/0503156}}.

\bibitem{Johnson:1999qt}
C.~V. Johnson, A.~W. Peet, and J.~Polchinski, \emph{{Gauge theory and the
  excision of repulson singularities}},
  \href{http://dx.doi.org/10.1103/PhysRevD.61.086001}{Phys.Rev. {\bf D61}
  (2000)  086001}, \href{http://arxiv.org/abs/hep-th/9911161}{{\tt
  arXiv:hep-th/9911161 [hep-th]}}.

\bibitem{Petrini:2001fk}
M.~Petrini, R.~Russo, and A.~Zaffaroni, \emph{{N=2 gauge theories and systems
  with fractional branes}},
  \href{http://dx.doi.org/10.1016/S0550-3213(01)00270-X}{Nucl.Phys. {\bf B608}
  (2001)  145--161}, \href{http://arxiv.org/abs/hep-th/0104026}{{\tt
  arXiv:hep-th/0104026 [hep-th]}}.

\bibitem{Nekrasov:2002qd}
N.~Nekrasov, \emph{{Seiberg-Witten prepotential from instanton counting}}, Adv.
  Theor. Math. Phys. {\bf 7} (2004)  831--864,
\href{http://arxiv.org/abs/hep-th/0206161}{{\tt arXiv:hep-th/0206161}}.

\bibitem{Nekrasov:2003rj}
N.~Nekrasov and A.~Okounkov, \emph{{Seiberg-Witten theory and random
  partitions}},
\href{http://arxiv.org/abs/hep-th/0306238}{{\tt arXiv:hep-th/0306238}}.

\bibitem{Green:2000ke}
M.~B. Green and M.~Gutperle, \emph{{D-instanton induced interactions on a
  D3-brane}}, JHEP {\bf 02} (2000)  014,
\href{http://arxiv.org/abs/hep-th/0002011}{{\tt arXiv:hep-th/0002011}}.

\bibitem{Billo:2002hm}
M.~Billo, M.~Frau, I.~Pesando, F.~Fucito, A.~Lerda, and A.~Liccardo,
  \emph{{Classical gauge instantons from open strings}}, JHEP {\bf 02} (2003)
  045,
\href{http://arxiv.org/abs/hep-th/0211250}{{\tt arXiv:hep-th/0211250}}.

\bibitem{Green:1997tv}
M.~B. Green and M.~Gutperle, \emph{{Effects of D-instantons}},
  \href{http://dx.doi.org/10.1016/S0550-3213(97)00269-1}{Nucl. Phys. {\bf B498}
  (1997)  195--227},
\href{http://arxiv.org/abs/hep-th/9701093}{{\tt arXiv:hep-th/9701093}}.

\bibitem{Green:1997tn}
M.~B. Green and M.~Gutperle, \emph{{D-particle bound states and the D-instanton
  measure}}, JHEP {\bf 01} (1998)  005,
\href{http://arxiv.org/abs/hep-th/9711107}{{\tt arXiv:hep-th/9711107}}.

\bibitem{Green:1998yf}
M.~B. Green and M.~Gutperle, \emph{{D-instanton partition functions}},
  \href{http://dx.doi.org/10.1103/PhysRevD.58.046007}{Phys. Rev. {\bf D58}
  (1998)  046007},
\href{http://arxiv.org/abs/hep-th/9804123}{{\tt arXiv:hep-th/9804123}}.

\bibitem{Witten:1995gx}
E.~Witten, \emph{{Small instantons in string theory}},
  \href{http://dx.doi.org/10.1016/0550-3213(95)00625-7}{Nucl. Phys. {\bf B460}
  (1996)  541--559},
\href{http://arxiv.org/abs/hep-th/9511030}{{\tt arXiv:hep-th/9511030}}.

\bibitem{Douglas:1995bn}
M.~R. Douglas, \emph{{Branes within branes}},
\href{http://arxiv.org/abs/hep-th/9512077}{{\tt arXiv:hep-th/9512077}}.

\bibitem{Moore:1998et}
G.~W. Moore, N.~Nekrasov, and S.~Shatashvili, \emph{{D-particle bound states
  and generalized instantons}},
  \href{http://dx.doi.org/10.1007/s002200050016}{Commun. Math. Phys. {\bf 209}
  (2000)  77--95},
\href{http://arxiv.org/abs/hep-th/9803265}{{\tt arXiv:hep-th/9803265}}.

\bibitem{Bruzzo:2002xf}
U.~Bruzzo, F.~Fucito, J.~F. Morales, and A.~Tanzini, \emph{{Multi-instanton
  calculus and equivariant cohomology}}, JHEP {\bf 05} (2003)  054,
\href{http://arxiv.org/abs/hep-th/0211108}{{\tt arXiv:hep-th/0211108}}.

\bibitem{Flume:2002az}
R.~Flume and R.~Poghossian, \emph{{An algorithm for the microscopic evaluation
  of the coefficients of the Seiberg-Witten prepotential}},
  \href{http://dx.doi.org/10.1142/S0217751X03013685}{Int. J. Mod. Phys. {\bf
  A18} (2003)  2541},
\href{http://arxiv.org/abs/hep-th/0208176}{{\tt arXiv:hep-th/0208176}}.

\bibitem{Billo:2009di}
M.~Billo, L.~Ferro, M.~Frau, L.~Gallot, A.~Lerda, and I.~Pesando, \emph{{Exotic
  instanton counting and heterotic/type I' duality}},
  \href{http://dx.doi.org/10.1088/1126-6708/2009/07/092}{JHEP {\bf 07} (2009)
  092},
\href{http://arxiv.org/abs/0905.4586}{{\tt arXiv:0905.4586 [hep-th]}}.

\bibitem{Fucito:2009rs}
F.~Fucito, J.~F. Morales, and R.~Poghossian, \emph{{Exotic prepotentials from
  D(-1)D7 dynamics}},
  \href{http://dx.doi.org/10.1088/1126-6708/2009/10/041}{JHEP {\bf 10} (2009)
  041},
\href{http://arxiv.org/abs/0906.3802}{{\tt arXiv:0906.3802 [hep-th]}}.

\bibitem{Billo':2010bd}
M.~Billo, M.~Frau, F.~Fucito, A.~Lerda, J.~F. Morales, and R.~Poghossian,
  \emph{{Stringy instanton corrections to N=2 gauge couplings}},
  \href{http://dx.doi.org/10.1007/JHEP05(2010)107}{JHEP {\bf 05} (2010)  107},
\href{http://arxiv.org/abs/1002.4322}{{\tt arXiv:1002.4322 [hep-th]}}.

\bibitem{Billo:2010mg}
M.~Billo, L.~Gallot, A.~Lerda, and I.~Pesando, \emph{{F-theoretic vs
  microscopic description of a conformal N=2 SYM theory}},
  \href{http://dx.doi.org/10.1007/JHEP11(2010)041}{JHEP {\bf 11} (2010)  041},
\href{http://arxiv.org/abs/1008.5240}{{\tt arXiv:1008.5240 [hep-th]}}.

\bibitem{Sen:1996vd}
A.~Sen, \emph{{F-theory and orientifolds}},
  \href{http://dx.doi.org/10.1016/0550-3213(96)00347-1}{Nucl. Phys. {\bf B475}
  (1996)  562--578},
\href{http://arxiv.org/abs/hep-th/9605150}{{\tt arXiv:hep-th/9605150}}.

\bibitem{Banks:1996nj}
T.~Banks, M.~R. Douglas, and N.~Seiberg, \emph{{Probing F-theory with branes}},
  \href{http://dx.doi.org/10.1016/0370-2693(96)00808-8}{Phys. Lett. {\bf B387}
  (1996)  278--281},
\href{http://arxiv.org/abs/hep-th/9605199}{{\tt arXiv:hep-th/9605199}}.

\bibitem{Schwarz:1983qr}
J.~H. Schwarz, \emph{{Covariant field equations of chiral N=2 D=10
  supergravity}},
  \href{http://dx.doi.org/10.1016/0550-3213(83)90192-X}{Nucl.Phys. {\bf B226}
  (1983)  269}.

\bibitem{Howe:1983sra}
P.~S. Howe and P.~C. West, \emph{{The complete N=2, D=10 supergravity}},
\href{http://dx.doi.org/10.1016/0550-3213(84)90472-3}{Nucl. Phys. {\bf B238}
  (1984)  181}.

\bibitem{deHaro:2002vk}
S.~de~Haro, A.~Sinkovics, and K.~Skenderis, \emph{{A supersymmetric completion
  of the R**4 term in IIB supergravity}},
  \href{http://dx.doi.org/10.1103/PhysRevD.67.084010}{Phys. Rev. {\bf D67}
  (2003)  084010},
\href{http://arxiv.org/abs/hep-th/0210080}{{\tt arXiv:hep-th/0210080}}.

\bibitem{Green:2003an}
M.~B. Green and C.~Stahn, \emph{{D3-branes on the Coulomb branch and
  instantons}}, JHEP {\bf 0309} (2003)  052,
  \href{http://arxiv.org/abs/hep-th/0308061}{{\tt arXiv:hep-th/0308061
  [hep-th]}}.

\bibitem{Polchinski:1995mt}
J.~Polchinski, \emph{{Dirichlet branes and Ramond-Ramond charges}},
  \href{http://dx.doi.org/10.1103/PhysRevLett.75.4724}{Phys.Rev.Lett. {\bf 75}
  (1995)  4724--4727}, \href{http://arxiv.org/abs/hep-th/9510017}{{\tt
  arXiv:hep-th/9510017 [hep-th]}}.

\bibitem{DiVecchia:1997pr}
P.~Di~Vecchia, M.~Frau, I.~Pesando, S.~Sciuto, A.~Lerda, and R.~Russo,
  \emph{{Classical p-branes from boundary state}},
  \href{http://dx.doi.org/10.1016/S0550-3213(97)00576-2}{Nucl. Phys. {\bf B507}
  (1997)  259--276},
\href{http://arxiv.org/abs/hep-th/9707068}{{\tt arXiv:hep-th/9707068}}.

\bibitem{Green:1987mn}
M.~B. Green, J.~H. Schwarz, and E.~Witten, \emph{{Superstring Theory. Vol.
  2}},. Cambridge Univ. Pr. ( 1987) ( Cambridge Monographs On Mathematical
  Physics).

\bibitem{Billo':2009gc}
M.~Billo, M.~Frau, L.~Gallot, A.~Lerda, and I.~Pesando, \emph{{Classical
  solutions for exotic instantons?}},
  \href{http://dx.doi.org/10.1088/1126-6708/2009/03/056}{JHEP {\bf 03} (2009)
  056},
\href{http://arxiv.org/abs/0901.1666}{{\tt arXiv:0901.1666 [hep-th]}}.

\bibitem{Grossman:1989bb}
B.~Grossman, T.~W. Kephart, and J.~D. Stasheff, \emph{{Solutions to gauge field
  equations in eight-dimensions: conformal invariance and the last Hopf map}},
\href{http://dx.doi.org/10.1016/0370-2693(89)90898-8}{Phys. Lett. {\bf B220}
  (1989)  431}.

\bibitem{Grossman:1984pi}
B.~Grossman, T.~W. Kephart, and J.~D. Stasheff, \emph{{Solutions to Yang-Mills
  field equations in eight-dimensions and the last Hopf map}},
\href{http://dx.doi.org/10.1007/BF01212529}{Commun. Math. Phys. {\bf 96} (1984)
   431 (Erratum: ibid. {\bf 100} (1985) 311)}.

\bibitem{Polchinski:1994fq}
J.~Polchinski, \emph{{Combinatorics of boundaries in string theory}},
  \href{http://dx.doi.org/10.1103/PhysRevD.50.R6041}{Phys. Rev. {\bf D50}
  (1994)  6041--6045},
\href{http://arxiv.org/abs/hep-th/9407031}{{\tt arXiv:hep-th/9407031}}.

\bibitem{Billo:2006jm}
M.~Billo, M.~Frau, F.~Fucito, and A.~Lerda, \emph{{Instanton calculus in R-R
  background and the topological string}}, JHEP {\bf 11} (2006)  012,
\href{http://arxiv.org/abs/hep-th/0606013}{{\tt arXiv:hep-th/0606013}}.

\bibitem{Ito:2010vx}
K.~Ito {\em et al.}, \emph{{N=2 instanton effective action in Omega-background
  and D3/D(-1) brane system in R-R background}},
  \href{http://dx.doi.org/10.1007/JHEP11(2010)093}{JHEP {\bf 11} (2010)  093},
\href{http://arxiv.org/abs/1009.1212}{{\tt arXiv:1009.1212 [hep-th]}}.

\end{thebibliography}

\end{document}